\documentclass[lettersize,journal]{IEEEtran}
\IEEEoverridecommandlockouts
\usepackage[utf8]{inputenc} 
\usepackage[T1]{fontenc}
\usepackage{url}
\usepackage{ifthen}
\usepackage{cite}
\usepackage[cmex10]{amsmath} 
\usepackage{amssymb}
\usepackage{graphicx}
\usepackage{epstopdf}
\usepackage{epsfig}
\usepackage{multirow}
\usepackage{amsthm}
\usepackage{comment}
\usepackage{caption}
\usepackage{enumitem}
\usepackage{amsmath}
\usepackage{color}
\usepackage{bbm}
\usepackage{algorithm}
\usepackage{algpseudocode}
\usepackage{csquotes}
\usepackage{subfiles}
\usepackage{caption, multirow, makecell}
\usepackage{array}
\usepackage{caption}
\usepackage{afterpage}
\usepackage{dblfloatfix}
\usepackage{hyperref}
\usepackage{graphicx}
\usepackage{balance}
\usepackage{subcaption}
\usepackage{diagbox}
\usepackage{xcolor}
\usepackage{float}
\newtheorem{thm}{Theorem}
\newtheorem{lem}{Lemma}

\newtheorem{cor}{Corollary}
\newtheorem{example}{Example}

\newtheorem{defn}{Definition}

\newtheorem{rem}{Remark}

\def\BibTeX{{\rm B\kern-.05em{\sc i\kern-.025em b}\kern-.08em
		T\kern-.1667em\lower.7ex\hbox{E}\kern-.125emX}}
\begin{document}
	
	\title{Hierarchical Coded Caching with Low Subpacketization and Coding Delay using Combinatorial $t$-Designs\\}
	
	\author{\IEEEauthorblockN{Rashid Ummer N.T., \textit{Member, IEEE} and B. Sundar Rajan, \textit{Life Fellow, IEEE}} }
	
	\maketitle
	
	\begin{abstract}
	Multi-layer cache structures are commonly used in various IoT environments due to their distributed architecture. This paper considers coded caching for a two-layer hierarchical network consisting of a single server connected to multiple cache-aided mirror sites, and each mirror site connected to a distinct set of cache-aided users. The placement delivery array (PDA) and the hierarchical placement delivery array (HPDA)  were proposed as tools for designing coded caching schemes with reduced subpacketization levels for single-layer and two-layer networks, respectively. This paper proposes construction of a novel class of HPDAs by first constructing a class of PDAs using combinatorial $t$-designs. The proposed class of HPDAs yields hierarchical coded caching schemes at several memory points for a given number of mirrors and users. Additionally, we introduce the concept of hierarchical memory sharing to achieve the lower envelope of the convex hull of these points. It is shown that the proposed hierarchical schemes have significantly lower subpacketization levels compared to many known schemes. In cases where the system parameters and subpacketization levels of the proposed and existing schemes match, the proposed scheme achieves a better coding delay. Furthermore, the class of PDAs constructed either subsumes several known PDA constructions or achieves a better transmission load for the same system parameters.
	\end{abstract}
	\section{Introduction}
	Wireless data traffic has grown substantially in recent years, driven predominantly by video-on-demand and the Internet of Things (IoT), and is projected to increase by 20–30 percent annually \cite{Eri}.  {\footnotetext{The authors are with the Department of Electrical Communication Engineering, Indian Institute of Science, Bangalore, 560012, India (e-mail:\{rashidummer,bsrajan\}@iisc.ac.in).}} With cache-aided communication, some of the peak-time data traffic can be shifted to off-peak times. Maddah-Ali and Niesen in \cite{MaN} proposed the novel coded caching scheme (referred to as MAN scheme) for an error-free broadcast network consisting of a single server with a library of $N$ files connected to a set of $K$ users, each having a cache of size $M$ files. Under uncoded placement, the MAN scheme achieves the optimal transmission load $R$ when $N \geq K$ \cite{YMA}. However, achieving this load requires splitting each file into $F = \binom{K}{\frac{KM}{N}}$ packets, referred to as the subpacketization level, which grows exponentially with $K$ for a given memory ratio $\frac{M}{N}$. For example, for $K=70$ with $\frac{M}{N}=\frac{1}{5}$, $F \approx 1.9 \times 10^{14}$. In this case, the file size has to be at least $22$ terabytes. Furthermore, any practical scheme requires header information for each subfile to enable decoding. Thus, low subpacketization level coded caching schemes are desirable for practical implementations. Yan \textit{et al.} in \cite{YCT} proposed an array called Placement Delivery Array (PDA) that describes both the placement and delivery phases of a coded caching scheme. Low subpacketization level schemes can be obtained by designing appropriate PDAs, and several such schemes have been proposed for single-layer networks.
	
	\begin{figure}[!htbp]
		\centering
		\captionsetup{justification=centering}
		\includegraphics[width=0.49\textwidth]{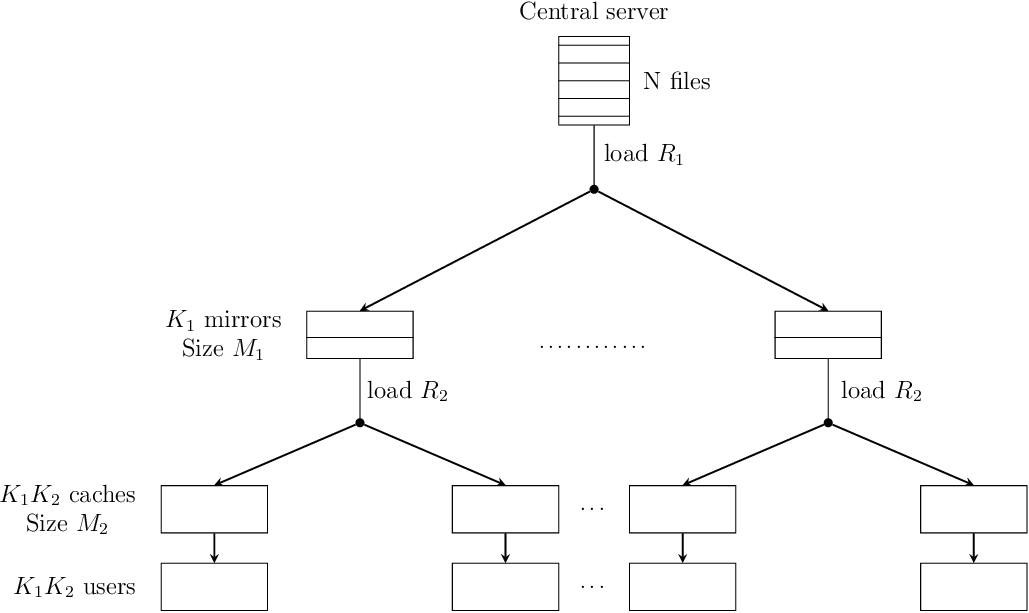}
		\caption{The two-layer $(K_1,K_2;M_1,M_2;N)$ hierarchical caching system.}
		\label{fig:setting2}
	\end{figure}
	Multi-layer cache structures are commonly employed in various IoT environments due to their distributed architecture. A hierarchical shared cache system for content distribution in industrial IoT (IIoT) applications, with hierarchically interconnected edge devices having cache memories, is considered in \cite{CB}. A cache-enabled hierarchical IIoT network with one macro base station connected to several small base stations, each serving a set of users, is studied in \cite{GTZZ}. A hierarchical edge caching architecture for the Internet of Vehicles is considered in  \cite{ZJHMW}.  Motivated by practical scenarios, Karamchandani \textit{et al.} in \cite{KNMD} proposed a coded caching scheme for a two-layer hierarchical network. A two-layer $(K_1,K_2;M_1,M_2;N)$ hierarchical caching system, depicted in Fig.\ref{fig:setting2}, consists of a server with a library of $N$ files, denoted by $\mathcal{W} = \{W_1, W_2, \dots, W_N\}$, connected through an error-free broadcast link to $K_1$ mirror sites, each having a cache of size $M_1$ files. Each mirror site is connected through an error-free broadcast link to $K_2$ users, each with a cache of size $M_2$ files. The loads from the server to the mirrors and from the mirrors to the users are denoted by $R_1$ and $R_2$, respectively.
		
	Coded caching schemes for a two-layer hierarchical network are also discussed in \cite{KYWM}, \cite{ZZWXL} and \cite{WWCY}. All the schemes in \cite{KNMD}, Theorem 2 in \cite{KYWM}, \cite{ZZWXL} and \cite{WWCY} utilize the MAN scheme and thus require a subpacketization level that grows exponentially with the number of users, which is undesirable for practical implementations. Kong \textit{et al.} in \cite{KYWM} proposed an array called hierarchical placement delivery array (HPDA) that describes both placement and delivery phases of a hierarchical coded caching scheme. Low subpacketization level hierarchical schemes can be obtained by constructing appropriate HPDAs. One class of HPDA construction in \cite{KYWM} using two given PDAs results in schemes with reduced subpacketization levels. No other existing work has discussed the subpacketization problem in hierarchical coded caching. All the schemes in \cite{KNMD}-\cite{WWCY} considered uncoded data placement and uniform user distribution.
	To compare different hierarchical coded caching schemes, the metric considered in \cite{WWCY} is \textit{coding delay} defined as the duration of the delivery phase normalized to the file size. The objective is to design low subpacketization level hierarchical schemes that minimize the coding delay. In this paper, we construct a novel class of HPDAs that achieves both low subpacketization level and low coding delay.
		
	The contributions of this paper are as follows: \\
	\noindent $\bullet$  We construct a new class of HPDAs with low subpacketization levels by first constructing a class of PDAs using combinatorial $t$-designs. The proposed class of HPDAs gives hierarchical coded caching schemes for a fixed $K_1$ and $K_2$ at several memory points.  We propose the concept of hierarchical memory sharing by which the lower envelope of the convex hull of the obtained memory-transmission load tuples $(M_1, M_2, R_1, R_2)$ is also achievable.\\
	\noindent $\bullet$  It is shown that the proposed hierarchical schemes have significantly lower subpacketization levels compared to many known hierarchical schemes. In cases where the system parameters and the subpacketization level of the proposed scheme and the existing scheme match, the proposed scheme has a better coding delay. \\
	\noindent $\bullet$  The class of PDAs constructed using $t$-designs either subsumes several known PDAs or gives coded caching schemes for a single-layer network that achieves better transmission load for the same number of users and memory ratio.
	
	The rest of the paper is organized as follows. In Section \ref{prelim}, we briefly review the preliminaries of PDA, HPDA and combinatorial $t$-design. The novel HPDA construction, by first constructing a class of PDAs using $t$-designs, is discussed in Section \ref{constr1}. In Section \ref{perform_anlysis1}, the performance analysis of the proposed class of HPDAs is carried out by comparing them with existing hierarchical coded caching schemes. In Section \ref{perform_anlysis2}, the performance analysis of the class of PDAs constructed in Section \ref{constr1} is carried out. Section \ref{concl} concludes the paper.
	
	\textit{Notations}: For any positive integer $n$, $[n]$ denotes the set $\{1,2,...,n\}$. For any integers $i$ and $n$ such that $0 \le i \le n$, $\binom{n}{i}$ denotes the binomial coefficient, which is calculated as $\frac{n!}{i!(n-i)!}$. For any set $\mathcal{A}$, $|\mathcal{A}|$ denotes the cardinality of $\mathcal{A}$. For a set $\mathcal{A}$ and a positive integer $i \leq |\mathcal{A}|$,  $\binom{\mathcal{A}}{i}$ denotes all the $i$-sized subsets of $\mathcal{A}$. For sets $\mathcal{A} \text{ and }\mathcal{B}$, $\mathcal{A}\textbackslash \mathcal{B}$ denotes the elements in $\mathcal{A}$ but not in $\mathcal{B}$. 
	
		\section{Preliminaries}\label{prelim}
		In this section, we first review the concepts of PDA and HPDA, and briefly describe the performance metric, coding delay. Then we review some basic definitions and properties from combinatorial designs that is used in this paper.	
	\subsection{Placement Delivery Array (PDA)}
	\begin{defn}
		(\hspace{1sp}\cite{YCT}) For positive integers $K, F, Z$ and $S$, an $F \times K$ array $\mathbf{P}=(p_{j,k})_{j \in [F], k \in [K]}$, composed of a specific symbol $\star$ and $S$ non-negative integers $1,2,...,S$, is called a $(K, F, Z, S)$ PDA if it satisfies the following conditions: \\
		\textit{C1}. The symbol $\star$ appears $Z$ times in each column.\\
		\textit{C2}. Each integer occurs at least once in the array.\\
		\textit{C3}. For any two distinct entries $p_{j_1,k_1}$ and $p_{j_2,k_2}$, we have $p_{j_1,k_1}=p_{j_2,k_2}=s$ is an integer only if (\textit{a}) $j_1 \neq j_2$, $k_1 \neq k_2$, i.e., they lie in distinct rows and distinct columns, and (\textit{b}) $p_{j_1,k_2}=p_{j_2,k_1}=\star$.
	\end{defn} 
	\begin{defn}
		(\hspace{1sp}\cite{YCT}) An array $\mathbf{P}$ is said to be a $g$-regular $(K, F, Z, S)$ PDA or $g-(K, F, Z, S)$ PDA, if it satisfies \textit{C1}, \textit{C3}, and the following condition \\
		\textit{C2'}. Each integer appears $g$ times in $\mathbf{P}$ where $g$ is a constant.
	\end{defn}
	\begin{lem}\label{lem:pda_scheme}
		(\hspace{1sp}\cite{YCT}) For a given $(K, F, Z, S)$ PDA $\mathbf{P}=(p_{j,k})_{F \times K}$, a $(K,M,N)$ coded caching scheme can be obtained with subpacketization $F$ and $\frac{M}{N}=\frac{Z}{F}$. Any possible demands from all users are met with a load of $R=\frac{S}{F}$.
	\end{lem}
	\subsection{Hierarchical Placement Delivery Array (HPDA)}
	\begin{defn}
		(\hspace{1sp}\cite{KYWM}) For any given positive integers $K_1, K_2, F, Z_1, Z_2$ with $Z_1 < F$, $Z_2 < F $ and any integer sets $S_m$ and $S_{k_1}, k_1 \in [K_1]$, an  $ F \times (K_1+K_1K_2)$ array $\mathbf{Q}= \left( \mathbf{Q^{(0)}},\mathbf{Q^{(1)}},....,\mathbf{Q^{(K_1)}} \right)$, where $\mathbf{Q^{(0)}}=(q_{j,k_1}^{(0)})_{j \in [F], k_1 \in [K_1]}$, is an $F \times K_1$ array consisting of $\star$ and \textit{null}  and $\mathbf{Q^{(k_1)}}=(q_{j,k_2}^{(k_1)})_{j \in [F], k_2 \in [K_2]}$ is an $F \times K_2$  array over  $\{ \star \cup S_{k_1} \}, k_1 \in [K_1]$, is a $(K_1, K_2;F;Z_1,Z_2;S_m,S_1,....,S_{K_1})$ hierarchical placement delivery array  if it satisfies the following conditions: \\
		\textit{B1}. Each column of $\mathbf{Q^{(0)}}$ has $Z_1$ stars.\\
		\textit{B2}. $\mathbf{Q^{(k_1)}}$ is a $(K_2,F,Z_2,|S_{k_1}|)$ PDA for each $ k_1 \in [K_1]$ .\\
		\textit{B3}. Each integer $s \in S_m$ occurs in exactly one subarray $\mathbf{Q^{(k_1)}}$ where $ k_1 \in [K_1]$ and for each $(q_{j,k_2}^{(k_1)})=s \in S_m,j \in [F],k_1 \in [K_1], k_2 \in [K_2]$, we have $(q_{j,k_1}^{(0)})=\star$. \\
		\textit{B4}. For any two entries $q_{j,k_2}^{(k_1)}$ and $q_{j',k'_2}^{(k'_1)}$, where $k_1 \neq k'_1 \in [K_1]$, $j,j' \in [F]$ and $k_2,k'_2 \in [K_2]$, if $q_{j,k_2}^{(k_1)}=q_{j',k'_2}^{(k'_1)}$ is an integer then
		\begin{enumerate}[label=-]		
			\item $q_{j',k_2}^{(k_1)}$ is an integer only if $q_{j',k_1}^{(0)}=\star$,
			\item $q_{j,k'_2}^{(k'_1)}$ is an integer only if $q_{j,k'_1}^{(0)}=\star$.\\
		\end{enumerate}
	\end{defn}  
	
	In a given HPDA $\mathbf{Q}= \left( \mathbf{Q^{(0)}},\mathbf{Q^{(1)}},....,\mathbf{Q^{(K_1)}} \right)$, $\mathbf{Q^{(0)}}$ indicates data placement in mirror sites and $\mathbf{Q^{(k_1)}}, k_1 \in [K_1]$ indicates placement at users attached to $k_1^{th}$ mirror site and delivery from the server and mirrors to these users. The set of users attached to the $k_1^{th}$ mirror site is denoted as $\mathcal{U}_{k_1}$ and the $k_2^{th}$ user in this set is denoted as $\mathcal{U}_{(k_1,k_2)}$. Cache contents of mirror site $k_1$ and user $\mathcal{U}_{(k_1,k_2)}$ are denoted as $\mathcal{Z}_{k_1}$ and $\mathcal{Z}_{(k_1,k_2)}$, respectively. The set of requested files is represented by a demand vector $\vec{d}=(d_{1,1},d_{1,2},\ldots,d_{K_1,K_2})$, where $d_{k_1,k_2}$ denotes the file demanded by user $\mathcal{U}_{(k_1,k_2)}$. The corresponding coded caching scheme is explained in Algorithm \ref{alg2}. 
	\vspace{-0.5cm}
	\begin{algorithm}[]
		\renewcommand{\thealgorithm}{1}
		\caption{\cite{KYWM} Caching scheme based on a HPDA $\mathbf{Q}$}
		\label{alg2}
		\begin{algorithmic}[1]
			\Procedure{Placement}{$\mathbf{Q},\mathcal{W}$}       
			\State Split each file $W_n \in \mathcal{W}$ into $F$ packets, i.e., $W_n =\{W_{n,j} | j \in [F]\}$
			\For{\texttt{$k_1 \in [K_1]$}}
			\State  $\mathcal{Z}_{k_1}$ $\leftarrow$ $\{W_{n,j}: q_{j,k_1}^{(0)}=\star, n \in [N],j \in [F]\}$
			\EndFor
			\For{\texttt{$(k_1,k_2), k_1 \in [K_1], k_2 \in [K_2] $}}
			\State $\mathcal{Z}_{(k_1,k_2)}$ $\leftarrow$ $\{W_{n,j}: q_{j,k_2}^{(k_1)}=\star, n \in [N],j \in [F]\}$
			\EndFor
			\EndProcedure		
			\Procedure{Delivery Server}{$\mathbf{Q},\mathcal{W},\vec{d} \hspace{0.1cm}$} 
			\For{\texttt{$s \in \left( \underset{k_1=1}{\bigcup^{K_1} S_{k_1}}\right) \textbackslash S_m $}}
			\State Server sends the following coded signal to the mirror sites:
			\State $X_s=\underset{q_{j,k_2}^{(k_1)}=s, j\in [F],k_1\in[K_1], k_2\in[K_2]}{\bigoplus}W_{d_{k_1,k_2},j}$
			\EndFor    
			\EndProcedure
			\Procedure{Delivery Mirrors}{$\mathbf{Q},\mathcal{W},\vec{d}, X_s $} 
			\For{\texttt{$k_1\in[K_1], s \in S_{k_1} \textbackslash S_m $}}
			\State After receiving $X_s$, mirror site $k_1$ sends the following coded signal to users in $\mathcal{U}_{k_1}$ :
			\State \hspace{-0.4cm} $X_{k_1,s}=X_s{\bigoplus}\left(\tiny \underset{\begin{array}{c} q_{j,k_2}^{(k'_1)}=s, q_{j,k_1}^{(0)}=\star, j\in [F], \\ k_2\in[K_2], k'_1\in[K_1]\textbackslash {k_1} \end{array}}{\bigoplus}W_{d_{k'_1,k_2},j}\right)$
			\EndFor
			\For{\texttt{$k_1\in[K_1], s' \in S_{k_1} \bigcap S_m$}}
			\State Mirror site $k_1$ sends the following coded signal to users in $\mathcal{U}_{k_1}$:
			\State $X_{k_1,s'}=\underset{q_{j,k_2}^{(k_1)}=s', j\in [F],k_2\in[K_2]}{\bigoplus}W_{d_{k_1,k_2},j}$
			\EndFor   
			\EndProcedure	
		\end{algorithmic}
	\end{algorithm}
	\vspace{-0.5cm}
	\begin{lem}\label{lem:hpda}
		(\hspace{1sp}\cite{KYWM}) Given a $(K_1,K_2;F;Z_1,Z_2;S_m,S_1,.., S_{K_1})$  HPDA $\mathbf{Q}= \left( \mathbf{Q^{(0)}},\mathbf{Q^{(1)}},....,\mathbf{Q^{(K_1)}} \right)$, we can obtain an $F$-division $(K_1,K_2;M_1,M_2;N)$ coded caching scheme using Algorithm \ref{alg2} with $\frac{M_1}{N}=\frac{Z_1}{F}$, $\frac{M_2}{N}=\frac{Z_2}{F}$ and transmission loads   $R_1=\frac{\left|\underset{k_1=1}{\bigcup^{K_1} S_{k_1}}\right|-\left|S_m \right|}{F}$  and  $R_2=\max_{k_1 \in [K_1]}\left\{ \frac{|S_{k_1}|}{F}\right\}$.
	\end{lem}	
	\noindent The following example illustrates Algorithm \ref{alg2} and Lemma \ref{lem:hpda}.
	\begin{example}
		It is easy to see that the following array $\mathbf{Q}$ is a $(3,2;6;2,3;S_m,S_1,S_2,S_3)$ HPDA where $S_m=\{4, 5, 6\}$, $S_1=\{1, 2, 4\}$, $S_2=\{1, 3, 5\}$ and $S_3=\{2, 3, 6\}$.\\
		Based on this HPDA $\mathbf{Q}$, one can obtain a $6-(3,2;2,3;6)$ hierarchical scheme using Algorithm \ref{alg2} as follows. \\
		\begin{equation*}
			\scalebox{0.85}{$			
			\mathbf{Q}	=	\begin{array}{cc|*{12}{c}}
				& & \multicolumn{3}{c}{\text{$\mathbf{Q^{(0)}}$}} & \vdots	& \multicolumn{2}{c}{\text{$\mathbf{Q^{(1)}}$}} & \vdots &\multicolumn{2}{c}{\text{$\mathbf{Q^{(2)}}$}} & \vdots  &\multicolumn{2}{c}{\text{$\mathbf{Q^{(3)}}$}} \\
				\hline
				k_1 & & 1 & 2 & 3  & \vdots	& \multicolumn{2}{c}{1} & \vdots &\multicolumn{2}{c}{2} & \vdots & \multicolumn{2}{c}{3}  \\
				\hline
				k_2 & & \multicolumn{3}{c}{\text{\phantom{0}}} & \vdots	& 1 & 2 &  \vdots & 1 & 2  & \vdots & 1 & 2     \\ 
				\hline
				&	1 & \star & \phantom{0} &\phantom{0} &\vdots& \star & 4 & \vdots & \star & 1 & \vdots & \star & 2 \\	
				&	2 & \star & \phantom{0} &\phantom{0} &\vdots& 4 & \star & \vdots & 1 & \star & \vdots & 2 & \star \\
				&	3 & \phantom{0} & \star &\phantom{0} &\vdots& \star & 1 & \vdots & \star & 5 & \vdots & \star & 3 \\
				F &	4 & \phantom{0} & \star &\phantom{0} &\vdots& 1 & \star & \vdots & 5 & \star & \vdots & 3 & \star \\
				&	5 & \phantom{0} & \phantom{0} &\star &\vdots& \star & 2 & \vdots & \star & 3 & \vdots & \star & 6 \\
				&	6 & \phantom{0} & \phantom{0} &\star &\vdots& 2 & \star & \vdots & 3 & \star & \vdots & 6 & \star \\
			\end{array}
		$}
		\end{equation*}	
			\noindent $\bullet$ \textbf{Placement phase:} From line $2$ of Algorithm \ref{alg2}, each file $W_n, \forall n \in [6]$ is split into $6$ packets, i.e., $W_n =\{W_{n,1},W_{n,2},W_{n,3},W_{n,4},W_{n,5},W_{n,6}\}$. By lines $3-5$ of Algorithm \ref{alg2}, the packets cached in each mirror site is given by, $\mathcal{Z}_1 = \{W_{n,1},W_{n,2}\}, \mathcal{Z}_2 = \{W_{n,3},W_{n,4}\}, \mathcal{Z}_3 = \{W_{n,5},W_{n,6}\}, \forall n \in [6]$. 
			By lines $6-8$ of Algorithm \ref{alg2}, each user caches the packets as follows:
			\begin{equation*}
				\begin{split}
					& \mathcal{Z}_{(1,1)} = \{W_{n,1},W_{n,3},W_{n,5}\}, \mathcal{Z}_{(1,2)} = \{W_{n,2},W_{n,4},W_{n,6}\}, \\ & \mathcal{Z}_{(2,1)} = \{W_{n,1},W_{n,3},W_{n,5}\}, \mathcal{Z}_{(2,2)} = \{W_{n,2},W_{n,4},W_{n,6}\}, \\ & \mathcal{Z}_{(3,1)} = \{W_{n,1},W_{n,3},W_{n,5}\}, \mathcal{Z}_{(3,2)} = \{W_{n,2},W_{n,4},W_{n,6}\}.
				\end{split}
			\end{equation*} \\
			\noindent $\bullet$ \textbf{Delivery phase}: Let the demand vector $\vec{d}=\{2,1,5,4,3,6\}$. By line $11$ of Algorithm \ref{alg2}, we have $ s \in \left( \underset{k_1=1}{\bigcup^{3} S_{k_1}}\right) \textbackslash S_m = \{ 1,2,3\}$. By lines $11-14$, server transmits to all mirror sites the following messages:
			$ X_1 = W_{2,4} \oplus W_{1,3} \oplus W_{5,2} \oplus W_{4,1}$, $X_2 = W_{2,6} \oplus W_{1,5} \oplus W_{3,2} \oplus W_{6,1}, X_3 = W_{5,6} \oplus W_{4,5} \oplus W_{3,4} \\ \oplus W_{6,3}$.
			Therefore the load of the first layer is $R_1=\frac{3}{6}=0.5$. \\ After receiving $X_1, X_2$ and $X_3$, by lines $17-20$ of Algorithm \ref{alg2}, the mirror site $k_1$ sends the messages $X_{k_1,s}, s \in  S_{k_1} \textbackslash S_m$, to its connected users. These messages are as follows: \\
			$X_{1,1} = X_1 \oplus W_{5,2} \oplus W_{4,1} = W_{2,4} \oplus W_{1,3}, X_{1,2} = W_{2,6} \oplus W_{1,5}, X_{2,1} = W_{5,2} \oplus W_{4,1}, X_{2,3} = W_{5,6} \oplus W_{4,5}, X_{3,2} = W_{3,2} \oplus W_{6,1}, X_{3,3} =  W_{3,4} \oplus W_{6,3}$.
			
			By lines $21-24$ of Algorithm \ref{alg2}, the mirror sites $1,2$ and $3$ send the messages $X_{1,4} = W_{2,2} \oplus W_{1,1}$, $X_{2,5} = W_{5,4} \oplus W_{4,3}$ and $X_{3,6} = W_{3,6} \oplus W_{6,5}$ respectively to its connected users. Therefore the load of the second layer is $R_2=\frac{2+1}{6}=0.5$.
			
			Each user can thus recover the requested file. For example, consider user $\mathcal{U}_{(2,1)}$. It has in its cache the packets $W_{5,1}, W_{5,3}$ and $W_{5,5}$ of the requested file $W_5$. From the transmissions $X_{2,1}$ and $X_{2,3}$, user $\mathcal{U}_{(2,1)}$ can recover $W_{5,2}$ and $W_{5,6}$ respectively, since it has $W_{4,1}$ and $W_{4,5}$. From the transmission $X_{2,5}$, user $\mathcal{U}_{(2,1)}$ can recover $W_{5,4}$ since it has $W_{4,3}$. Thus user $\mathcal{U}_{(2,1)}$ gets all the $6$ packets of the requested file $W_5$.
	\end{example}

	In a hierarchical coded caching network, a memory-transmission load tuple $(M_1, M_2, R_1, R_2)$ is said to be achievable if, for any possible demand vector $\vec{d}$, every user can successfully recover their desired file. The time to transmit one file size ($B$ bits) from the server to the mirrors and from the mirrors to its attached users are assumed to be one unit of time.  Then we define the performance metric, \textit{coding delay}, as follows.
	\begin{defn}
		For an achievable memory-load tuple $(M_1, M_2, R_1, R_2)$, let $T_d$ denote the duration of the delivery phase. The coding delay, $T$, is defined as the duration of the delivery phase normalized to the file size. i.e., $ T \triangleq \frac{T_d}{B}$. 
	\end{defn}
	Since the loads $R_1$ and $R_2$ are denoted in terms of files, if the mirror transmission occurs after the server finishes the transmission, then $T=R_1+R_2$. On the other hand, if parallel transmission between the two layers is possible, then the server and the mirrors concurrently transmit symbols throughout the transmission slots, and thus, $T=\max\{R_1, R_2\}$. Since $T$ is a function of $R_1$ and $R_2$, an achievable memory-load tuple $(M_1, M_2, R_1, R_2)$ can be equivalently expressed as an achievable memory-coding delay tuple $(M_1, M_2, T)$.
	
	In \cite{RCB} and \cite{JZF}, the performance metric considered for a two-layer hierarchical coded caching scheme is the \textit{composite rate}, defined as $\bar{R}=R_1+K_1R_2$, which, according to \cite{RCB}, quantifies the total system bandwidth. However, in most practical hierarchical networks, each mirror's broadcast messages are received only by its connected users, not by users connected to other mirrors. Thus, the same frequency bands can be reused across mirrors, making $R_1+R_2$, not $\bar{R}$, the correct metric quantifying the total system bandwidth. Furthermore, even when frequency reuse is not possible, the metric of coding delay remains significant as it directly reflects overall system latency.

	\subsection{Combinatorial t-design}
	Combinatorial $t$-designs is an important topic in design theory. The existence and constructions of large classes of $t$-designs are discussed in \cite{Stin}, \cite{Col} and \cite{DT}. 
	\begin{defn}[Design $(\mathcal{X}, \mathcal{A})$]
		A design is a pair  $(\mathcal{X}, \mathcal{A})$ such that the following properties are satisfied: \\
		\textit{1}. $\mathcal{X}$ is a set of elements called points, and\\
		\textit{2}. $\mathcal{A}$ is a collection (i.e., multiset) of nonempty subsets of $\mathcal{X}$ called blocks.
	\end{defn}
	\begin{defn}[$t-(v, k, \lambda)$ design]\label{def_tdes}
		Let $v,k,\lambda$ and $t$ be positive integers such that $v>k\geq t$. A $t-(v, k, \lambda)$ design is a design $(\mathcal{X}, \mathcal{A})$ such that the following properties are satisfied: \\
		\textit{1}. $|\mathcal{X}|=v$, \\
		\textit{2}. each block contains exactly $k$ points, and \\
		\textit{3}. every set of $t$ distinct points is contained in exactly $\lambda$ blocks.
	\end{defn}
	
	The general term $t$-design is used to indicate any $t-(v, k, \lambda)$ design. For notation simplicity, we write blocks of $t$-designs and its subsets in the form $abcd$ instead of $\{ a,b,c,d\}$. 
	\begin{example}\label{ex:example1}
		$\mathcal{X}=\{1,2,3,4,5,6,7\}$, $ \mathcal{A}=\{127,145,136,467,256,357,234\}$ is a $2-(7,3,1)$ design. 
	\end{example}
	The following Lemma will be used in the construction of PDAs subsequently.
	\begin{lem}\label{lem:Stin}
		(\hspace{1sp}\cite{Stin}) Suppose that $(\mathcal{X}, \mathcal{A})$ is a $t-(v, k, \lambda)$ design. Suppose that $\mathcal{Y} \subseteq \mathcal{X}$, where $|\mathcal{Y}|=s\leq t.$ Then there are exactly $\lambda_{s}=\frac{\lambda \binom{v-s}{t-s}}{\binom{k-s}{t-s}}$ blocks in $\mathcal{A}$ that contains all the points in $\mathcal{Y}$. Thus the number of blocks in $\mathcal{A}$, $b=\lambda_{0}=\frac{\lambda \binom{v}{t}}{\binom{k}{t}}$.
	\end{lem}
\section{Proposed HPDA Construction}\label{constr1}	
In this section, we construct a novel class of HPDAs that give hierarchical coded caching schemes with low subpacketization levels, by first constructing a class of PDAs using combinatorial $t$-designs. The proposed class of HPDAs provides hierarchical coded caching schemes with a fixed number of mirrors and users at multiple memory points. In Section \ref{subsec_VA}, we propose the concept of hierarchical memory sharing, by which schemes can be obtained at any memory point lying on the plane defined by three achievable memory points, for a given number of mirrors and users.

Lemma \ref{lem:pda_scheme} and Lemma \ref{lem:hpda} in Section \ref{prelim}, only provide schemes from a given PDA and HPDA, respectively, and do not describe a method for constructing PDAs and HPDAs. Thus, designing coded caching and hierarchical coded caching schemes with low subpacketization levels can be formulated as constructing appropriate PDAs and HPDAs, respectively. The only known HPDA construction that results in a hierarchical scheme with low subpacketization is presented in \cite{KYWM}. Therefore, the construction of novel classes of HPDAs is both of theoretical interest and practical relevance.  

 We use combinatorial $t$-designs to construct the novel class of HPDAs. The idea is to represent the mirrors in the hierarchical network by the blocks of a given $t$-design, while the users connected to each mirror are represented by the subsets of the corresponding $t$-design block. Property \textit{2} in Definition \ref{def_tdes} ensures that the number of users attached to each mirror is the same. The \textit{balance} property of $t$-designs (Property \textit{3} in Definition \ref{def_tdes}) makes them particularly suitable for designing PDAs and HPDAs. In a $t$-$(v,k,\lambda)$ design, the number of blocks $b$ typically satisfies $b \ge v$. This allows us to represent file packets by subsets of the points of the $t$-design, thereby resulting in a lower subpacketization level. By leveraging these properties of $t$-designs, we construct a class of PDAs and HPDAs as given in Theorem \ref{thm:pda} and Theorem \ref{thm:HPDA}, respectively. 

\begin{figure}[!htbp]
	\begin{center}
		\captionsetup{justification=centering}
		\includegraphics[width=0.5\textwidth]{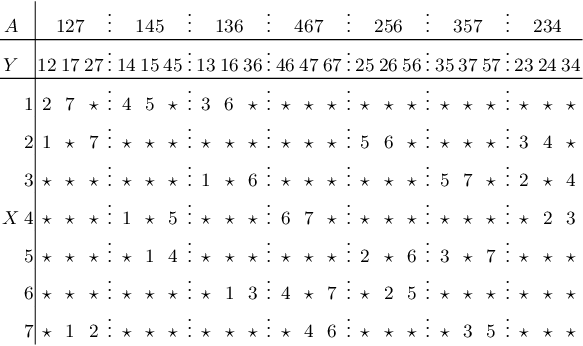}
		\caption{ $(21,7,5,7)$ PDA }
		\label{fig:example2}
	\end{center}
\end{figure}

\begin{thm}\label{thm:pda}
	Given a $t-(v,k,\lambda)$ design, for any positive integers $i$ and $j$ such that $i \le j \le t$, there exists a $\binom{v-j+i}{i}-\left( \frac{\lambda\binom{v}{t}\binom{k}{j}}{\binom{k}{t}},\binom{v}{i},\binom{v}{i}-\binom{j}{i},\frac{\binom{v}{j-i}\lambda\binom{v-j}{t-j}}{\binom{k-j}{t-j}} \right)$ regular PDA which gives a $\left(\frac{\lambda\binom{v}{t}\binom{k}{j}}{\binom{k}{t}}, M, N \right)$ coded caching scheme with subpacketization level $F=\binom{v}{i}$, memory ratio $\frac{M}{N}=1-\frac{\binom{j}{i}}{\binom{v}{i}}$ and transmission load $R=\frac{\binom{v}{j-i}\lambda\binom{v-j}{t-j}}{\binom{k-j}{t-j}\binom{v}{i}}$.
\end{thm}
\begin{IEEEproof}
	Let $(\mathcal{X}, \mathcal{A})$ be a $t-(v,k,\lambda)$ design. For any positive integers $i$ and $j$ such that $i \le j \le t$, define an array $\mathbf{P}$  whose rows are indexed by $X \in \binom{\mathcal{X}}{i}$  and columns by all the elements of $C=\{ (A,Y) : A \in \mathcal{A}, Y \in \binom{A}{j}\}$, as follows.
	\begin{equation}\label{eq:cons2}
		P_{X,(A,Y)}= \begin{cases}  (Y\textbackslash X)_\alpha & \text {if } X \subseteq Y   \\ \hspace{0.5 cm} \star & \text {otherwise, }\end{cases}
	\end{equation}
	where the subscript $\alpha$ denotes the $\alpha^{th}$ occurrence of $Y\textbackslash X$ from left to right in the row indexed by $X$. The subpacketization level $F$ is the number of rows of $\mathbf{P}$, which is equal to $\binom{v}{i}$. The number of users $K$ is given by the number of columns of $\mathbf{P}$. Since columns are indexed by $C=\{ (A,Y) : A \in \mathcal{A}, Y=\binom{A}{j}\}$, the number of columns is the number of blocks in $\mathcal{A}$ times the number of $j$ sized subsets of $k$ sized blocks. Therefore,  $K=\frac{\lambda\binom{v}{t}}{\binom{k}{t}}\binom{k}{j}$. 
	
	The non $\star$ entries of $\mathbf{P}$ are denoted by $(Y\textbackslash X)_\alpha$ and $|Y\textbackslash X|=(j-i)$. Therefore, the number of possible $Y\textbackslash X$ is $\binom{v}{j-i}$. To find the number of occurrences of $Y\textbackslash X$ in a given row $X$, by $(\ref{eq:cons2})$, we need to find the number of blocks in which $\{Y\textbackslash X\} \cup X = Y$ belongs. By Lemma \ref{lem:Stin}, that is equal to $\frac{\lambda\binom{v-j}{t-j}}{\binom{k-j}{t-j}}$. Therefore, 
	$\alpha \in \left[1,\frac{\lambda\binom{v-j}{t-j}}{\binom{k-j}{t-j}}\right].$ 
	Thus $S= \frac{\binom{v}{j-i}\lambda\binom{v-j}{t-j}}{\binom{k-j}{t-j}}$.
		
	Let all the $(j-i)$ sized subsets of $\mathcal{X}$ are  arranged in some order, then we define a bijection $f(.)$ from $\left\{ \binom{\mathcal{X}}{j-i} \right\}$ to $\binom{v}{j-i}$. Thus, we can replace all the non-star entries of $\mathbf{P}$,	$(Y\textbackslash X)_\alpha$, by $(\alpha-1)\binom{v}{j-i}+f\left(Y\textbackslash X\right)$. This gives an array $\mathbf{P}$ with $S$ distinct non-negative integers. Now we have to show that the array $\mathbf{P}$ is the PDA specified in Theorem \ref{thm:pda}, which is done in \textit{Appendix \ref{appendix:PDA_proof}}.
	By Lemma \ref{lem:pda_scheme}, the obtained $\binom{v-j+i}{i}-\left( \frac{\lambda\binom{v}{t}\binom{k}{j}}{\binom{k}{t}},\binom{v}{i},\binom{v}{i}-\binom{j}{i},\frac{\binom{v}{j-i}\lambda\binom{v-j}{t-j}}{\binom{k-j}{t-j}} \right)$ regular PDA gives a $\left(\frac{\lambda\binom{v}{t}\binom{k}{j}}{\binom{k}{t}}, M, N \right)$ coded caching scheme with subpacketization level $F=\binom{v}{i}$, memory ratio $\frac{M}{N}=\frac{Z}{F}=1-\frac{\binom{j}{i}}{\binom{v}{i}}$ and transmission load $R=\frac{S}{F}=\frac{\binom{v}{j-i}\lambda\binom{v-j}{t-j}}{\binom{k-j}{t-j}\binom{v}{i}}$.
\end{IEEEproof} 
The following example illustrates Theorem \ref{thm:pda}.
\begin{example}
	Consider the $2-(7,3,1)$ design $(\mathcal{X}, \mathcal{A})$ in Example $\ref{ex:example1}$, $j=2$ and $i=1$. The columns are indexed by the elements of $C=\{ (A,Y) : A \in \mathcal{A}, Y \in \binom{A}{2}\}$. There are $7$ blocks each of size $3$ in the given design. Therefore, the number of users $K= 7 \times \binom{3}{2} =21$. The rows are indexed by the points in $\mathcal{X}$, since $i=1$. Therefore the subpacketization level is $7$. Now, using (\ref{eq:cons2}), we obtain the $(21,7,5,7)$ PDA shown in Fig. \ref{fig:example2}. To make clear how the entries are obtained, consider the entry, $P_{1,(136,16)}$. Since $ 1 \subset 16$, by (\ref{eq:cons2}), $P_{1,(136,16)}=(16\textbackslash 1)_\alpha = 6_\alpha$. $\alpha$ denotes the $\alpha^{th}$ occurrence of $6$ from left to right in the row $X=1$. Since $1 \cup 6 =16$ occurs exactly in $1$ block, $6$ appears only once in the row $X=1$. Therefore, $\alpha \in [1]$.  The subscript $\alpha$ is omitted in Fig. \ref{fig:example2} since $\alpha$ can take value only $1$ in this example. From this PDA, one can obtain a coded caching scheme for $K=21$ and $\frac{M}{N}=\frac{5}{7}$ which requires a subpacketization level of $7$ and achieves a load $R=\frac{S}{F}=1$. 
\end{example} 
\begin{figure*}[hb]
	\begin{center}
		\captionsetup{justification=centering}
		\includegraphics[width=0.95\textwidth]{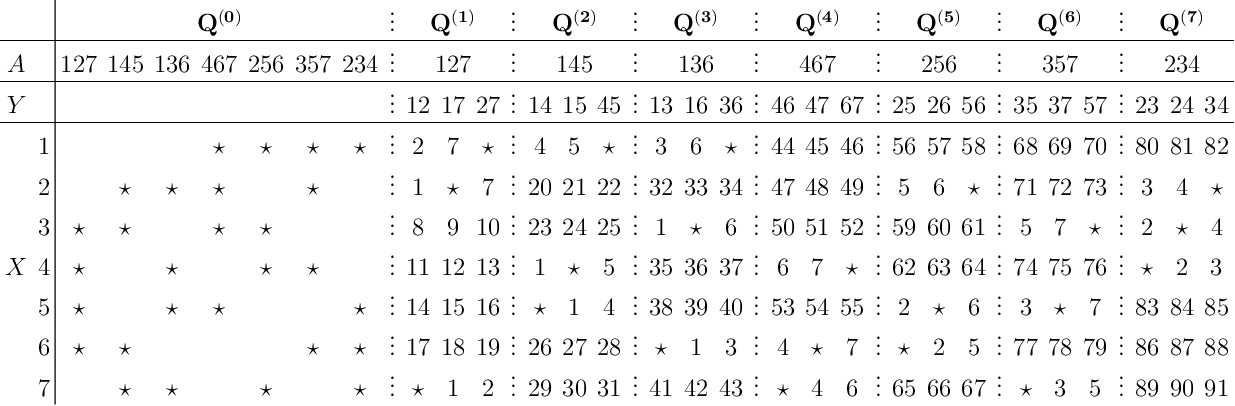}
		\caption{ Example of HPDA in Theorem \ref{thm:HPDA}}
		\label{fig:HPDA1}
	\end{center}
\end{figure*}
From the class of PDAs given by Theorem \ref{thm:pda}, we now obtain a new class of HPDAs given by Theorem \ref{thm:HPDA}. 
		\begin{thm}\label{thm:HPDA}
		Given a $t-(v,k,\lambda)$ design, for any positive integers $i$ and $j$ such that $i \le j \le t$, there exist a $\left(\frac{\lambda\binom{v}{t}}{\binom{k}{t}},\binom{k}{j};\binom{v}{i}; \right. $ $\left. \binom{v}{i}-\binom{k}{i},\binom{k}{i}-\binom{j}{i};S_m, S_1,..,S_{K_1} \right)$ HPDA which gives an $\binom{v}{i}$-division $\left(\frac{\lambda\binom{v}{t}}{\binom{k}{t}},\binom{k}{j};M_1, M_2; N \right)$ coded caching scheme with memory ratios $\frac{M_1}{N}=1-\frac{\binom{k}{i}}{\binom{v}{i}}$, $\frac{M_2}{N}=\frac{\binom{k}{i}-\binom{j}{i}}{\binom{v}{i}}$  and transmission loads $R_1=\frac{\binom{v}{j-i}\lambda\binom{v-j}{t-j}}{\binom{k-j}{t-j}\binom{v}{i}}$ and $R_2 \leq \frac{\frac{\binom{k}{j-i}\lambda\binom{v-j}{t-j}}{\binom{k-j}{t-j}}+\left[\binom{v}{i}-\binom{k}{i}\right]\binom{k}{j}}{\binom{v}{i}}$.
	\end{thm} 
	\begin{IEEEproof} 
		The proof follows from the HPDA construction described below. 
		\subsubsection{HPDA Construction}  The HPDA in Theorem \ref{thm:HPDA} can be constructed from the PDA $\mathbf{P}$ constructed in Theorem \ref{thm:pda} from $t$-design. Let each block $A \in \mathcal{A}$ in the $t$-design represent a mirror site. Then the PDA $\mathbf{P}$, constructed by replacing the non star entries of array in $(\ref{eq:cons2})$ by $(\alpha-1)\binom{v}{j-i}+f\left(Y\textbackslash X\right)$, can be partitioned into $K_1=|\mathcal{A}|=\frac{\lambda\binom{v}{t}}{\binom{k}{t}}$ parts by column. Let $\mathbf{P}=\left(\mathbf{P^{(1)}},\mathbf{P^{(2)}},....,\mathbf{P^{(K_1)}} \right)$. 
	\begin{itemize}
			\item Construction of $\mathbf{Q^{(0)}}$. 
			A row of any given array is called a \textit{star row} if this row contains only $\star$ entries. We construct the $F \times K_1$ mirror placement array $\mathbf{Q^{(0)}}=(q_{X,k_1}^{(0)})_{X \in \binom{\mathcal{X}}{i}, k_1 \in [K_1]}$ where
			\begin{equation}\label{eq:Mirror}
				q_{X,k_1}^{(0)}= \begin{cases}  \star & \text { if } P_{X,(A,Y)}=\star \hspace{0.2cm} \forall \hspace{0.1cm} Y \subset A \\ null & \text { otherwise. }\end{cases}
			\end{equation}
			i.e., $q_{X,k_1}^{(0)}$ is a $ \star $ if the row $X$ of $\mathbf{P^{(k_1)}}$ is a star row.
			\item Construction of $\mathbf{Q^{(k_1)}}$.
			Since each $A \in \mathcal{A}$ represent a mirror site, the columns corresponding to a given $A$ represent the number of attached users. Since $|A|=k$ and $Y \in \binom{A}{j}$, from $(\ref{eq:cons2})$, the number of users attached to each mirror site, $K_2=\binom{k}{j}$. Then $\mathbf{Q^{(k_1)}}$ is constructed by replacing the star entries in the star rows of $\mathbf{P^{(k_1)}}$ by distinct integers which has no intersection with $[S]$. 
			\item Construction of $\mathbf{Q}$.
			We get an $F \times (K_1+K_1K_2)$ array by arranging $ \mathbf{Q^{(0)}}$ and $\mathbf{Q^{(1)}},....,\mathbf{Q^{(K_1)}}$ horizontally. i.e., $\mathbf{Q}= \left( \mathbf{Q^{(0)}},\mathbf{Q^{(1)}},....,\mathbf{Q^{(K_1)}} \right)$.
		\end{itemize}
		
		Now we have to verify that the array $\mathbf{Q}$ satisfies the definition of HPDA. That is shown in \textit{Appendix \ref{appendix:hpda}}. Using Lemma \ref{lem:hpda} and Algorithm \ref{alg2}, the obtained $\left(\frac{\lambda\binom{v}{t}}{\binom{k}{t}},\binom{k}{j};\binom{v}{i}; \right. $ $\left. \binom{v}{i}-\binom{k}{i},\binom{k}{i}-\binom{j}{i};S_m, S_1,..,S_{K_1} \right)$ HPDA gives an $\binom{v}{i}$ division $\left(\frac{\lambda\binom{v}{t}}{\binom{k}{t}},\binom{k}{j};M_1, M_2; N \right)$ coded caching scheme with memory ratios $\frac{M_1}{N}=1-\frac{\binom{k}{i}}{\binom{v}{i}}$, $\frac{M_2}{N}=\frac{\binom{k}{i}-\binom{j}{i}}{\binom{v}{i}}$  and loads $R_1=\frac{\left|\underset{k_1=1}{\bigcup^{K_1} S_{k_1}}\right|-\left|S_m \right|}{F} = \frac{S}{F}= \frac{\binom{v}{j-i}\lambda\binom{v-j}{t-j}}{\binom{k-j}{t-j}\binom{v}{i}}$ and \\ $R_2=\max_{k_1 \in [K_1]}\left\{ \frac{|S_{k_1}|}{F}\right\} \leq \frac{\frac{\binom{k}{j-i}\lambda\binom{v-j}{t-j}}{\binom{k-j}{t-j}}+\left[\binom{v}{i}-\binom{k}{i}\right]\binom{k}{j}}{\binom{v}{i}}$.
		\end{IEEEproof}
		
	The following example illustrates Theorem \ref{thm:HPDA}.
\begin{example}
	From the $(21,7,5,7)$ PDA constructed using $2-(7,3,1)$ design for $j=2$ and $i=1$ shown in Fig. \ref{fig:example2}, we obtain a $(7, 3; 7; 4,1;S_m, S_1,..,S_7)$ HPDA shown in Fig. \ref{fig:HPDA1}. The mirrors are represented by the blocks of the $2-(7,3,1)$ design. Therefore, there are $7$ mirrors and the $3$ subsets of a given block represent the $3$ users attached to that mirror. The $7 \times 7$ mirror placement array $\mathbf{Q^{(0)}}$ is constructed by using (\ref{eq:Mirror}). For example, consider the columns corresponding to the block $127$ in the $(21,7,5,7)$ PDA in Fig. \ref{fig:example2}, denoted as $\mathbf{P^{(1)}}$. There are $4$ star rows in $\mathbf{P^{(1)}}$, in the rows $X=3,4,5$ and $6$. Therefore, we put stars in these rows in the column denoted by the block $127$ in the mirror placement array. Therefore, $Z_1=4$. Then $\mathbf{Q^{(1)}}$ is constructed by replacing the star entries in these star rows of $\mathbf{P^{(1)}}$ by distinct integers. Therefore, $Z_2=5-4=1$. The set of integers $\underset{k_1=1}{\bigcup^{K_1} S_{k_1}}\textbackslash S_m$ is same as the set of integers $[S]=[7]$ in the $(21,7,5,7)$ PDA. These integers correspond to the transmission from the server to the mirrors. The integers in each sub-array $\mathbf{Q^{(k_1)}}$, $S_{k_1}$, correspond to the transmission from the $k_1^{th}$ mirror to its attached users. From Fig. \ref{fig:HPDA1} it is clear that $|S_{k_1}|=15$, $\forall k_1 \in [7]$.  From the obtained HPDA in Fig. \ref{fig:HPDA1}, using Algorithm \ref{alg2} and Lemma \ref{lem:hpda}, we get hierarchical coded caching scheme with $\frac{M_1}{N}=\frac{4}{7}$, $\frac{M_2}{N}=\frac{1}{7}$, $F=7$ and transmission loads $R_1=\frac{S}{F}=1$ and $R_2=\frac{15}{7}=2.14$. The coding delay is $T=R_1+R_2=3.14$.  
\end{example} 
\begin{figure*}[!hbp]
	\centering
	\begin{subfigure}[b]{0.85\textwidth}
		\centering
		\captionsetup{justification=centering}
		\includegraphics[width=0.85\textwidth]{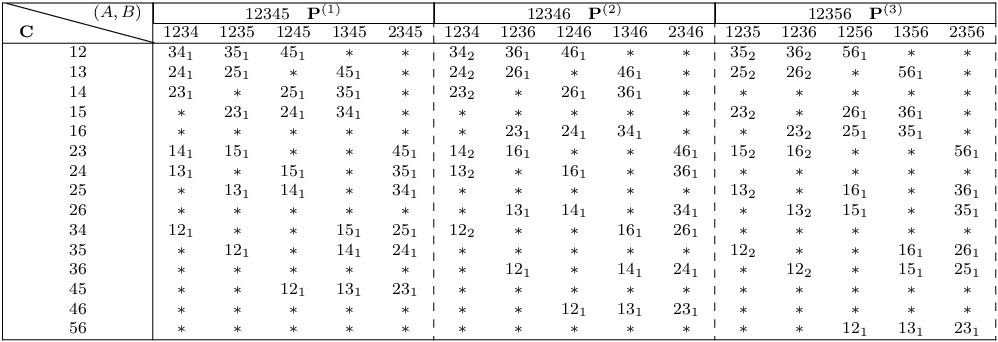}
		\caption{Column $1-15$ }
	\end{subfigure}
	\hfill
	\begin{subfigure}[b]{0.85\textwidth}
		\centering
		\captionsetup{justification=centering}
		\includegraphics[width=0.85\textwidth]{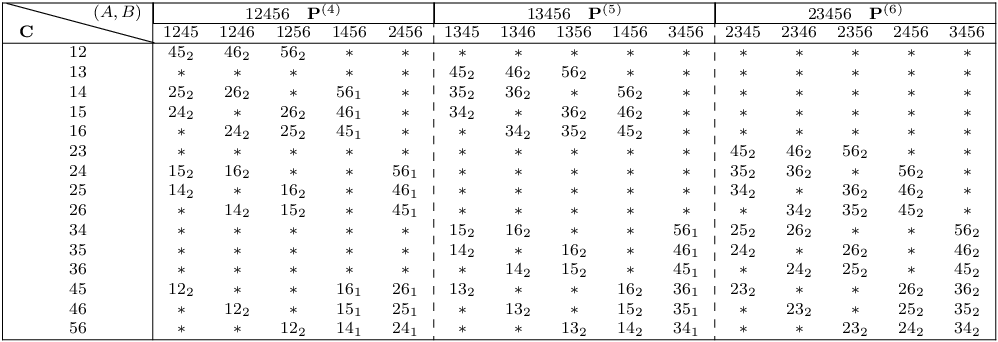}
		\caption{Column $16-30$ }
	\end{subfigure}
	\caption{$15\times 30$ array obtained by Corollary \ref{cor:HPDA} for $n=6, k=5, j=4$ and $i=2$}
	\label{fig:pda}
\end{figure*}
 \begin{figure*}[ht]
 	\centering
 	\begin{subfigure}[b]{0.865\textwidth}
 		\centering
 		\captionsetup{justification=centering}
 		\includegraphics[width=0.85\textwidth]{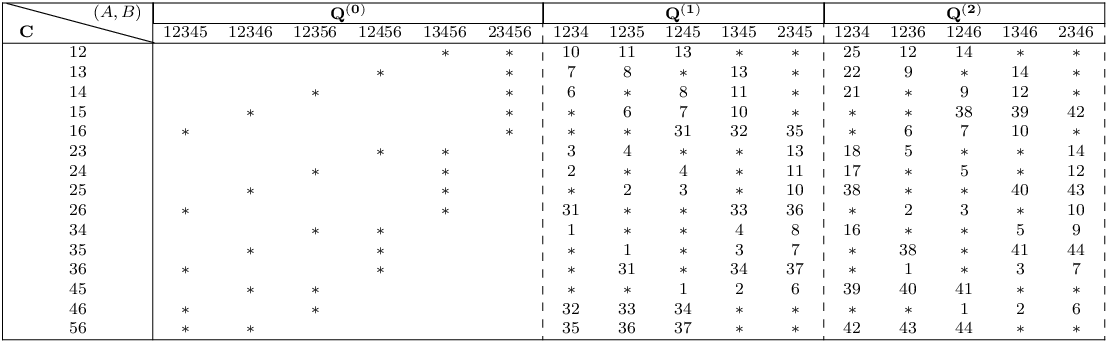}
 		\caption{$\mathbf{Q^{(0)}},\mathbf{Q^{(1)}},\mathbf{Q^{(2)}}$ of the HPDA $\mathbf{Q}= \left( \mathbf{Q^{(0)}},\mathbf{Q^{(1)}},....,\mathbf{Q^{(6)}} \right)$ }
 	\end{subfigure}
 	\hfill
 	\begin{subfigure}[b]{0.95\textwidth}
 		\centering
 		\captionsetup{justification=centering}
 		\includegraphics[width=0.975\textwidth]{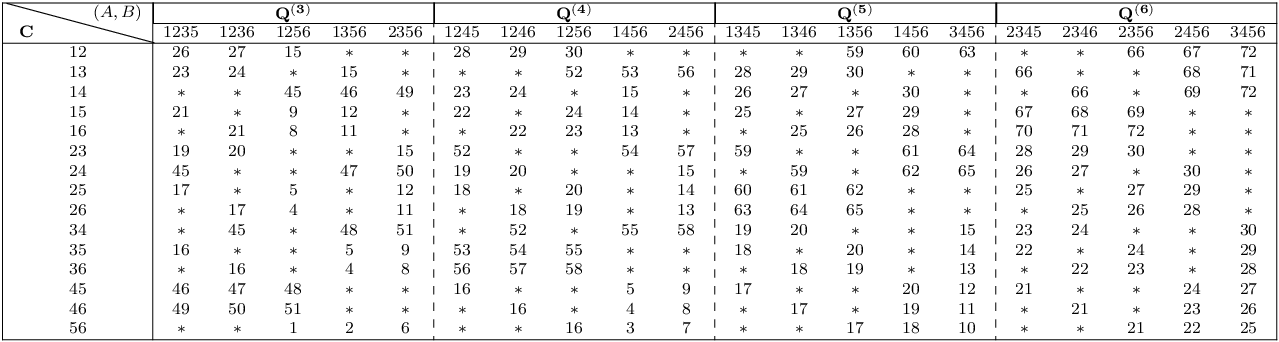}
 		\caption{$\mathbf{Q^{(3)}},\mathbf{Q^{(4)}},\mathbf{Q^{(5)}},\mathbf{Q^{(6)}}$ of the HPDA $\mathbf{Q}= \left( \mathbf{Q^{(0)}},\mathbf{Q^{(1)}},....,\mathbf{Q^{(6)}} \right)$ }
 	\end{subfigure}
 	\caption{HPDA obtained for $n=6, k=5, j=4$ and $i=2$ and $(5,5,2,7)$ PDA in Corollary \ref{cor:hpda}. }
 	\label{fig:HPDA2}
 \end{figure*}
From Theorem \ref{thm:pda} and Theorem \ref{thm:HPDA}, we obtain the following corollary, which enables the construction of a wide range of PDAs and HPDAs.
\begin{cor}\label{cor:HPDA}
		For any positive integers $i,j,k \text{ and } n$ such that $i \leq j \leq k \leq n$, there exists
		\begin{itemize}
			\item a $\binom{n-j+i}{i}-\left( \binom{n}{k}\binom{k}{j},\binom{n}{i},\binom{n}{i}-\binom{j}{i},  \binom{n}{j-i}\binom{n-j}{k-j} \right)$ PDA which gives a $\left(\binom{n}{k}\binom{k}{j},M,N \right)$ coded caching scheme with subpacketization level $\binom{n}{i}$, memory ratio $\frac{M}{N}=1-\frac{\binom{j}{i}}{\binom{n}{i}}$ and transmission load $R=\frac{\binom{n}{j-i}\binom{n-j}{k-j}}{\binom{n}{i}}$,
			\item a 
			$\left(\binom{n}{k},\binom{k}{j};\binom{n}{i};\binom{n}{i}-\binom{k}{i},\binom{k}{i}-\binom{j}{i};  S_m,S_1,..,S_{K_1} \right)$ HPDA which gives an $\binom{n}{i}$-division $\left(\binom{n}{k},\binom{k}{j};M_1,M_2;N \right)$ hierarchical coded caching scheme with memory ratios $\frac{M_1}{N}=1-\frac{\binom{k}{i}}{\binom{n}{i}}$, $\frac{M_2}{N}=\frac{\binom{k}{i}-\binom{j}{i}}{\binom{n}{i}}$  and transmission loads $R_1=\frac{\binom{n}{j-i}\binom{n-j}{k-j}}{\binom{n}{i}}$ and $R_2 \leq \frac{\binom{k}{j-i}\binom{n-j}{k-j}+\binom{k}{j} \left[ \binom{n}{i}-\binom{k}{i} \right] }{\binom{n}{i}}$.
		\end{itemize} 
\end{cor}
\begin{IEEEproof}
	For any positive integers $k$ and $n$ such that $k \le n$, taking every $k$-subset of $[n]$ results in a trivial $k-\left(n,k,1\right)$ design $(\mathcal{X}, \mathcal{A})$, where $\mathcal{X}=[n]$ and $\mathcal{A}=\binom{[n]}{k}$.  Using this design, Theorem \ref{thm:pda} gives a $\binom{n-j+i}{i}-\left( \binom{n}{k}\binom{k}{j},\binom{n}{i},\binom{n}{i}-\binom{j}{i},\binom{n}{j-i}\binom{n-j}{k-j} \right)$ regular PDA, where $i$, $j$, $k$ and $n$ are positive integers such that $i \le j \le k \le n$. Using Lemma \ref{lem:pda_scheme}, this obtained PDA gives the coded caching scheme specified in Corollary \ref{cor:HPDA}. Using the same $k-\left(n,k,1\right)$ design, Theorem \ref{thm:HPDA} gives a 
	$\left(\binom{n}{k},\binom{k}{j};\binom{n}{i};\binom{n}{i}-\binom{k}{i},\binom{k}{i}-\binom{j}{i};  S_m,S_1,..,S_{K_1} \right)$ HPDA.  Using Lemma \ref{lem:hpda} and Algorithm \ref{alg2}, this obtained HPDA gives an $\binom{n}{i}$-division $\left(\binom{n}{k},\binom{k}{j};M_1,M_2; N \right)$ hierarchical coded caching scheme with $\frac{M_1}{N}=1-\frac{\binom{k}{i}}{\binom{n}{i}}$, $\frac{M_2}{N}=\frac{\binom{k}{i}-\binom{j}{i}}{\binom{n}{i}}$  and $R_1=\frac{\binom{n}{j-i}\binom{n-j}{k-j}}{\binom{n}{i}}$ and $R_2 \leq \frac{\binom{k}{j-i}\binom{n-j}{k-j}+\binom{k}{j} \left[ \binom{n}{i}-\binom{k}{i} \right] }{\binom{n}{i}}$. 
\end{IEEEproof}
\begin{rem}
	The PDA in Corollary $1$ of \cite{Li} is a $\binom{n-j+i}{i}-\left(\binom{n}{i},\binom{n}{j},\binom{n}{j}-\binom{n-i}{j-i},\binom{n}{j-i}\right)$ PDA, for positive integers $i,j$ and $n$ such that $i \le j < n$. The transpose of this PDA is a $\binom{n-j+i}{i}-\left(\binom{n}{j},\binom{n}{i},\binom{n}{i}-\binom{j}{i},\binom{n}{j-i}\right)$ PDA. By concatenating this PDA horizontally $\binom{n-j}{k-j}$ times, where $j < k \le n$, with distinct set of integers in each, the PDA in Corollary \ref{cor:HPDA} can be obtained. However, the grouping of $\binom{n}{k}\binom{k}{j}$ users into $\binom{n}{k}$ sets, each containing  $\binom{k}{j}$ users, to obtain the HPDA in Corollary \ref{cor:HPDA} is not evident from such a concatenated construction. 
\end{rem} 

The following corollary provides hierarchical coded caching schemes at different user cache memory points for the same system parameters $K_1$, $K_2$ and $\frac{M_1}{N}$ in Theorem \ref{thm:HPDA} while preserving the subpacketization level and first-layer load.
\begin{cor}\label{cor:hpda}
	In the construction of HPDAs in Theorem \ref{thm:HPDA}, $\mathbf{Q^{(k_1)}}$ is constructed by replacing the star entries in the star rows of $\mathbf{P^{(k_1)}}$ by distinct integers which has no intersection with $[S]$. Instead of that replace the star rows of $\mathbf{P^{(k_1)}}$ by a PDA, $\mathbf{P'^{(k_1)}}= \left(K_2,Z_1,Z',S'\right)$ $\equiv$ $\left(\binom{k}{j},\binom{v}{i}-\binom{k}{i},Z',S'\right)$, where the $S'$ integers in each of $P^{'k_1}$ are distinct integers which has no intersection with $[S]$. This will result in a $\left(\binom{v}{k},\binom{k}{j};\binom{v}{i};\binom{v}{i}-\binom{k}{i},\binom{k}{i}-\binom{j}{i}+Z';S_m,S_1,..,S_{K_1} \right)$ HPDA which gives an $\binom{v}{i}$-division $\left(\binom{n}{k},\binom{k}{j};M_1,M_2;N \right)$ coded caching scheme with memory ratios $\frac{M_1}{N}=1-\frac{\binom{k}{i}}{\binom{v}{i}}$, $\frac{M_2}{N}=\frac{\binom{k}{i}-\binom{j}{i}+Z'}{\binom{v}{k}}$  and transmission load $R_1=\frac{\binom{v}{j-i}\binom{v-j}{k-j}}{\binom{v}{i}}$ and $R_2 \leq \frac{\binom{k}{j-i}\binom{v-j}{k-j}+S' }{\binom{v}{k}}$, for any positive integers $j \text{ and } i$ such that $i\leq j \leq t$. $Z'=0$ results in HPDA in Theorem \ref{thm:HPDA}.
\end{cor}
\begin{IEEEproof}
	The proof follows from the proof of Theorem \ref{thm:HPDA} except that now there are $Z'$ additional stars in every column of $\mathbf{Q^{(k_1)}}$ and $S^{''}_{k1}=\left[S+1+(k_1-1)S': S + k_1S'\right]$.
\end{IEEEproof}
 By choosing $\left(K_2,Z_1,Z',S'\right)$ PDA in Corollary \ref{cor:hpda} with different values of $Z' \in [Z_2]$, one can obtain hierarchical coded caching schemes at different user cache memory points $M_2$ for a given $K_1, K_2$ and $M_1$, while maintaining the same subpacketization level $F$ and transmission load $R_1$. The following example illustrates Corollary \ref{cor:HPDA} and  Corollary \ref{cor:hpda}.
\begin{example}\label{ex:pda55}
For $n=6$ and $ k=5$, $\mathcal{X}=\{1,2,3,4,5,6\}$, $\mathcal{A}=\{12345,12346,12356,12456,13456,23456\}$ is a $5-(6,5,1)$ design. Using this design, for $j=4$ and $i=2$, (\ref{eq:cons2}) gives a $15\times 30$ array shown in Fig.\ref{fig:pda}. Consider a bijection $f(.)$ from $\binom{[6]}{2}$ to $\left[15\right]$. Replacing the non-$\star$ entries $X_\alpha$ in the obtained array in Fig.\ref{fig:pda} by $(\alpha-1)15+f(X)$, where $X \in \binom{[6]}{2}$ and $\alpha \in [2]$, results in a $6-(30,15,9,30)$ PDA, which is the PDA specified in Corollary \ref{cor:HPDA}. It is clear from Fig.\ref{fig:pda} that in each $\mathbf{P}^{(k_1)}$, where $k_1 \in [6]$, there are $5$ star rows. Therefore, by Corollary \ref{cor:hpda}, using a $(5,5,2,7)$ Optimal \footnote{A $(K,F,Z,S)$ PDA is said to be an \textit{optimal PDA} if the number of distinct integers $S$ is the least possible integer such that a $(K,F,Z,S)$ PDA exists.} PDA 
$
		\mathbf{P} = \begin{bmatrix}
			\star & \star & 1 & 2 & 5 \\
			1 & \star & \star & 3 & 6 \\
			\star & 1 & \star & 4 & 7 \\
			2 & 3 & 4 & \star & \star \\
			5 & 6 & 7 & \star & \star \\
		\end{bmatrix}
$, we obtain a $(6, 5; 15; 5,6;S_m, S_1,..,S_6)$ HPDA shown in Fig.\ref{fig:HPDA2}, which gives an $15$-division $(6,5;M_1,M_2;N)$ coded caching scheme with $\frac{M_1}{N}=\frac{1}{3}$, $\frac{M_2}{N}=\frac{2}{5}$, $R_1=2$ and $R_2=1.8$. In this example, using different $(5,5,Z',S')$ PDAs instead of the $(5,5,2,7)$ PDA results in hierarchical schemes with $K_1=6$, $K_2=5$, $F=15$, $\frac{M_1}{N}=\frac{1}{3}$ and $R_1=2$, for different user cache memory ratios $\frac{M_2}{N}$ as shown in Table \ref{tab:HPDAs}.
\end{example}
\vspace{-0.45cm}
\begin{table}[H]
	\centering
	\begin{tabular}{| c | c | c |}
		\hline
		\rule{0pt}{4ex}
		\makecell{$(5,5,Z',S')$ Optimal PDAs  in \cite{Wei}}  & $\frac{M_2}{N}$ & $R_2$\\  	
		\hline
		\rule{0pt}{3.5ex}
		$(5,5,1,10)$ PDA   & $\frac{5}{15}$ & $2$\\
		\hline
		\rule{0pt}{3.5ex}
		$(5,5,3,4)$ PDA  & $\frac{7}{15}$ & $1.6$ \\
		\hline
		\rule{0pt}{3.5ex}
		$(5,5,4,1)$ PDA  & $\frac{8}{15}$ & $1.4$\\
		\hline
	\end{tabular} 
	\centering
	\caption{$\frac{M_2}{N}$ and $R_2$ for different PDAs in Example \ref{ex:pda55}.}
	\label{tab:HPDAs}
\end{table}
By the HPDA construction given in Theorem \ref{thm:HPDA}, for a given $t-(v,k,\lambda)$ design and positive integer $j \le t$, varying $i$ from $1 \text{ to } j$ will give hierarchical coded caching schemes for a fixed $K_1$ and $K_2$ at $j$ different memory points $(M_1,M_2)$. $(M_1,N,0)$ and $(0,0,N+K_2)$ are two trivial achievable points. Now we propose the concept of hierarchical memory sharing by which the lower envelope of the convex hull of the obtained points is also achievable.
\subsection{Hierarchical memory sharing}\label{subsec_VA}
\begin{figure}[!htbp]
	\centering
	\captionsetup{justification=centering}
	\includegraphics[width=0.45\textwidth]{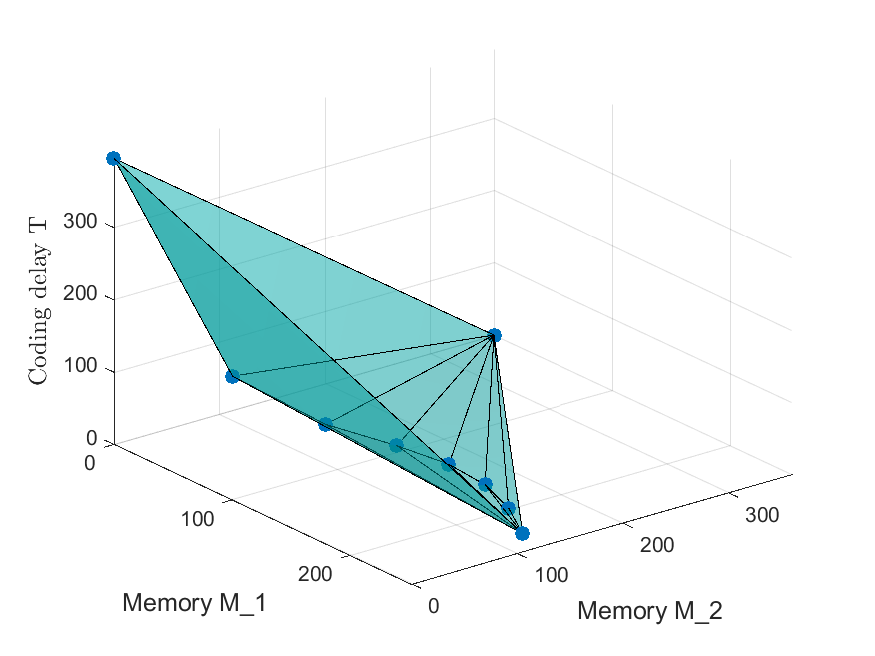}
	\caption{The coding delay as a function of $(M_1,M_2)$ for $K_1=10$, $K_2=36$ and $N=360$.}
	\label{fig:hier_mem}
\end{figure}

Let $(M'_1,M'_2,R'_1,R'_2)$, $(M''_1,M''_2,R''_1,R''_2)$ and $(M'''_1,M'''_2,\allowbreak R'''_1,R'''_2)$ be three achievable memory-load tuples for a given $K_1$ and $K_2$. These tuples can be equivalently denoted as $(M'_1, M'_2, T')$, $(M''_1, M''_2, T'')$ and $(M'''_1, M'''_2, T''')$ where $T$ is the coding delay. Any memory point $(M_1, M_2)$ that lie on the plane joining the memory points $(M'_1, M'_2)$, $(M''_1, M''_2)$ and $(M'''_1, M'''_2)$ can be written as a convex linear combination of this three points. i.e., $(M_1, M_2)=\alpha(M'_1, M'_2)+\beta (M''_1, M''_2)+(1-\alpha-\beta)(M'''_1, M'''_2) \text{ for } 0\leq \alpha, \beta \leq 1$. Then by memory sharing between these three achievable points,  any  $(M_1, M_2, T)$ tuple that lie on the triangular plane joining these three points is achievable. We proceed as follows.  Divide all the files $W_n : n \in N$ (each of size $B$ bits) into three parts $W_n^{(1)}$, $W_n^{(2)}$ and $W_n^{(3)}$  having $B_1=\alpha B$ bits, $B_2=\beta B$ bits and $B_3=(1-\alpha-\beta) B$ bits respectively, where $0 \leq \alpha, \beta \leq 1$. Then corresponding to  $W_n^{(1)} : n \in N$ we use the $(M'_1, M'_2, T')$ achievable scheme, corresponding to $W_n^{(2)} : n \in N$ we use the $(M''_1, M''_2, T'')$ achievable scheme and corresponding to $W_n^{(3)} : n \in N$ we use the $(M'''_1, M'''_2, T''')$ achievable scheme. Let $T$ be the corresponding coding delay. Then, $TB= T'B_1+T''B_2+T'''B_3 = T'\alpha B+T''\beta B+T'''(1-\alpha-\beta) B$. That is, $T=\alpha T' +\beta T'' +(1-\alpha-\beta) T'''$. That implies the lower convex envelope of the three points  $(M'_1, M'_2, T')$, $(M''_1, M''_2, T'')$ and $(M'''_1, M'''_2, T''')$ is achievable. This we refer to as \textit{hierarchical memory sharing}.
For example, for $n=10, k=9 \text{ and } j=7$, Corollary \ref{cor:HPDA} give schemes with $K_1=10 \text{ and } K_2=36$ for 7 memory points shown in Table \ref{tab:HPDA}. By hierarchical memory sharing, the lower envelope of the convex hull of these points are achievable, as shown in Fig.\ref{fig:hier_mem}.
\begin{table}[H]
	\centering
	\begin{tabular}{| c | c | c | c | c | c | c | }
		\hline
		\rule{0pt}{3.5ex}
		$i$  & $F$ & $\frac{M_1}{N}$ & $\frac{M_2}{N}$ &  $R_1$ & $R_2$ & $T=R_1+R_2$\\ [4pt] 	
		\hline
		\rule{0pt}{3.5ex}
		1 & $10$ & $\frac{1}{10}$& $\frac{1}{5}$ & $63.00$ & $28.80$ & $91.80$\\	
		\hline
		\rule{0pt}{3.5ex}
		2 & $45$ & $\frac{1}{5}$ & $\frac{1}{3}$ & $16.80$ & $15.60$ & $32.40$ \\
		\hline
		\rule{0pt}{3.5ex}
		3  & $120$ & $\frac{3}{10}$ & $\frac{49}{120}$ & $5.25$ & $13.95$ & $19.20$ \\
		\hline
		\rule{0pt}{3.5ex}
		4 & $210$ & $\frac{2}{5}$ & $\frac{13}{30}$ & $1.714$ & $15.60$ & $17.314$ \\
		\hline
		\rule{0pt}{3.5ex}
		5 & $252$ & $\frac{1}{2}$ & $\frac{5}{12}$ & $0.53$ & $18.43$ & $18.96$ \\
		\hline
		\rule{0pt}{3.5ex}
		6 & $210$ & $\frac{3}{5}$ & $\frac{11}{30}$ & $0.143$ & $21.73$ & $21.873$ \\
		\hline
		\rule{0pt}{3.5ex}
		7  & $120$ & $\frac{7}{10}$ & $\frac{7}{24}$ & $0.025$ & $25.22$ & $25.245$ \\
		\hline
	\end{tabular}
	\centering
	\caption{Hierarchical schemes by Corollary \ref{cor:HPDA} for  \\$n=10, k=9, j=7$ for different $i$ values.}
	\label{tab:HPDA}
\end{table}
\begin{rem}
	All the achievable points $(M_1^i, M_2^i, T^i) \text{ } \forall i \in [j]$ from Theorem \ref{thm:HPDA} will be meaningful, if they form a convex set, i.e., they lie on the lower envelope of the convex hull formed by these points. It is hard to show analytically that they form a convex set. Simulations results have showed that the achievable points form a convex set, as evident from Fig.\ref{fig:hier_mem}.
\end{rem}
\section{Performance analysis of the proposed class of HPDAs}\label{perform_anlysis1}
In this section, we compare the proposed class of HPDAs with the existing hierarchical coded caching schemes in \cite{KNMD}, \cite{KYWM}, \cite{ZZWXL} and \cite{WWCY}. The comparison is carried out by considering subpacketization level and coding delay as the metrics, for the same system parameters $K_1$, $K_2$, $\frac{M_1}{N}$ and $\frac{M_2}{N}$. For comparison, $R_2$ of proposed schemes is taken as the upper bound for $R_2$ given in the corresponding theorem. Except for the scheme in \cite{WWCY} which allows a concurrent transmission between the two layers, the coding delay is taken as $R_1+R_2$. 	
\subsection{Comparison with  \cite{KNMD}:} In \cite{KNMD}, a decentralized hierarchical coded caching scheme is proposed (referred to as the KNMD scheme), which combines two subschemes controlled by the parameters $\alpha, \beta \in [ 0,1 ]$. The first subscheme (Scheme A in \cite{KNMD}) delivers $\alpha$ fraction of the requested file to all the users by considering the entire mirror cache and $\beta$ fraction of each user's cache. The second subscheme (Scheme B in \cite{KNMD}) delivers the remaining $(1-\alpha)$ fraction of requested file to all the users by considering $(1-\beta)$ fraction of each user's cache and ignoring the mirror cache. The resulting transmission loads are given as 
\begin{equation}\label{eq:knmd}
	\footnotesize
	\begin{split}
		& R_1(\alpha,\beta) \triangleq   \alpha\cdot K_2\cdot r\left(\frac{M_1}{\alpha N},K_1\right)
		+(1-\alpha)\cdot r\left(\frac{(1-\beta)M_2}{(1-\alpha)N},K_1K_2 \right)  \\ & \text{\& } R_2(\alpha,\beta) \triangleq  \alpha\cdot r\left(\frac{\beta M_2}{\alpha N},K_2\right)
		+(1-\alpha)\cdot r\left(\frac{(1-\beta)M_2}{(1-\alpha)N},K_2 \right), 
	\end{split}
\end{equation}
	 for $\alpha, \beta \in [0,1]$. We consider this scheme with centralized data placement in \cite{MaN}, in which case $r\left(\frac{M}{N},K\right)  \triangleq K (1-M/N)  min \left\{ \frac{1}{1+KM/N}, \frac{N}{K} \right\} $, where $M_i\in\{0,{N}/{K_i},{2N}/{K_i},\ldots,N\}$, $i=1,2$. The centralized scheme achieving a load $r\left(\frac{M}{N},K\right)$ requires a subpacketization level of $\binom{K}{\frac{KM}{N}}$, where $\frac{KM}{N}$ is an integer. For comparison, the subpacketization level of the schemes in \cite{KNMD} is taken as the maximum subpacketization required in these schemes.  For $M_i\notin\{0,{N}/{K_i},{2N}/{K_i}, \ldots,N\}$, schemes can be obtained by memory sharing \footnote{The load obtained by memory sharing is $r\left(\frac{M}{N},K\right)  = \theta r\left(\frac{\lfloor \frac{KM}{N} \rfloor}{K},K\right)+ (1-\theta) r\left(\frac{\lceil \frac{KM}{N} \rceil}{K},K\right)$, where $\frac{KM}{N}=\theta \lfloor \frac{KM}{N} \rfloor + (1-\theta) \lceil \frac{KM}{N} \rceil$, for $0 < \theta < 1$ and subpacketization level $F=max \left\{ \binom{K}{\lfloor \frac{KM}{N} \rfloor } , \binom{K}{\lceil \frac{KM}{N} \rceil }\right\}$.}. The optimized values of $\alpha$ and $\beta$ with respect to both $R_1(\alpha,\beta)$ and $R_2(\alpha,\beta)$ is given in \cite{KNMD} by,
\begin{equation}\label{eq:knmd2}
	\footnotesize
	\scalebox{0.875}{$
	(\alpha^\star, \beta^\star) \triangleq
	\begin{cases}
		\left(\displaystyle \frac{M_1}{N}, \frac{M_1}{N} \right) 
		& \text{ if } M_1+M_2K_2 \geq N \mbox{ and } \\ & 0 \leq M_1 \leq N/4, \\[-0.5ex]
		\left( \displaystyle \frac{M_1}{M_1 + M_2K_2} , 0 \right)
		& \text{ if } M_1+M_2K_2 < N, \\[-0.3ex]
		\left(\displaystyle \frac{M_1}{N}, \frac{1}{4} \right)
		& \text{ if } M_1+M_2K_2 \geq N \mbox{ and }\\ &
		 N/4 < M_1 \leq N.
	\end{cases}
	$} 
\end{equation}

It can be verified that, for the system parameters of the proposed hierarchical coded caching schemes in Theorem \ref{thm:HPDA} and Corollary \ref{cor:HPDA}, $M_1 + K_2M_2 \ge N$. However, by using the concept of hierarchical memory sharing proposed in Section \ref{subsec_VA}, by choosing the trivial memory-load tuple $(M'_1,M'_2,R'_1,R'_2)=(0,0,N,K_2)$ and two memory-load tuples $(M''_1,M''_2,R''_1,R''_2)$ and $(M'''_1,M'''_2, R'''_1,R'''_2)$ obtained from the proposed schemes, one can obtain hierarchical coded caching schemes in the memory regime $M_1 + K_2M_2 < N$. This is illustrated with an example in row $9$ of Table \ref{tab:comp_HPDA}. 

\begin{table*}[!htbp]
	\centering
	\setlength{\tabcolsep}{3pt}
	\scriptsize
	\begin{tabular}{|c|c| c | c | c | c | c | c | c | c | c |}
		\hline
		\rule{0pt}{4ex}
		\makecell{Sl. \\ No.}&\makecell{Comparing Schemes}& \makecell{Parameters}  & \makecell{$K_1$} & \makecell{$K_2$}& \makecell{$\frac{M_1}{N}$} & \makecell{$\frac{M_2}{N}$} & \makecell{$F$}& \makecell{$R_1$} &\makecell{$R_2$}& \makecell{$T=$ \\ $R_1+R_2$} \\ [4pt] 	
		\hline
		\rule{0pt}{3.5ex}
		1 & \makecell{Proposed scheme \\ in Theorem \ref{thm:HPDA}}  & \makecell{for $3-(8,4,1)$ design \\ with $j=3$ and $i=2$} & $14$ & $4$ & $\frac{11}{14}$& $\frac{3}{28}$ & $28$ & $0.2857$ & $3.285$ & $3.5714$ \\
		\hline
		\rule{0pt}{3.5ex}
		2 & \makecell{KNMD scheme with\\ centralized data placement} & ($\alpha^{\star}=\frac{11}{14}, \beta^{\star}=\frac{1}{4}$)  & $14$ & $4$ & $\frac{11}{14}$& $\frac{3}{28}$ & $1.346\times10^{15}$ & $0.3409$ & $3.107$ & $3.448$ \\
		\hline
		\rule{0pt}{3.5ex}
		3 &\makecell{Scheme A in \cite{KNMD} with\\ centralized data placement}  & ($\alpha=1, \beta=1$)  & $14$ & $4$ & $\frac{11}{14}$& $\frac{3}{28}$ & $364$ & $1$ & $2.9285$ & $3.9285$ \\
		\hline
		\rule{0pt}{3.5ex}
		4 & \makecell{Scheme B in \cite{KNMD} with\\ centralized data placement} & ($\alpha=0, \beta=0$)  & $14$ & $4$ & $\frac{11}{14}$& $\frac{3}{28}$ & $3.247\times10^{7}$ & $7.1428$ & $2.9285$ & $10.0713$ \\
		\hline
		\hline 
		\rule{0pt}{3.5ex}
		5 &\makecell{Proposed scheme \\ in Corollary \ref{cor:HPDA}}  & for $n=7, k=6, j=5 \text{ and } i=1$ & $7$ & $6$& $\frac{1}{7}$ & $\frac{1}{7}$ & $7$ & $10$ & $5.14285$ & $15.14285$ \\	
		\hline
		\rule{0pt}{3.5ex}
		6 & \makecell{KNMD scheme with\\ centralized data placement} & ($\alpha^{\star}=\frac{1}{7}, \beta^{\star}=\frac{1}{7}$) & $7$ & $6$ & $\frac{1}{7}$ & $\frac{1}{7}$ & $5.24\times 10^{6}$ & $4.408$ & $3$ & $7.408$ \\
		\hline
		\rule{0pt}{3.5ex}
		7 & \makecell{Scheme A in \cite{KNMD} with\\ centralized data placement} & ($\alpha=1, \beta=1$) & $7$ & $6$ & $\frac{1}{7}$ & $\frac{1}{7}$ & $7$ & $18$ & $3$ & $21$ \\
		\hline
		\rule{0pt}{3.5ex}
		8 & \makecell{Scheme B in \cite{KNMD} with\\ centralized data placement} & ($\alpha=0, \beta=0$) & $7$ & $6$ & $\frac{1}{7}$ & $\frac{1}{7}$ & $5.24\times 10^{6}$ & $5.14285$ & $3$ & $8.14285$ \\
		\hline
		\hline 
		\rule{0pt}{3.5ex}
		9 &\makecell{Proposed scheme in \\ Corollary \ref{cor:HPDA} and \\ hierarchical memory sharing}  & \makecell{Hierarchical memory sharing using  the memory-load tuples- \\ i) $(0,0,42,6)$ (trivial), ii) $(6,6,10,5.14285)$ \\ (Corollary \ref{cor:HPDA}  for $n=7, k=6, j=5 \text{ and } i=1$) and \\ iii) $(12,10,3.333,3.619)$ (Corollary \ref{cor:HPDA} \\ for $n=7, k=6, j=5 \text{ and } i=2$), $\alpha=0.6$, $\beta=0.3$. } & $7$ & $6$& $\frac{1}{14}$ & $\frac{1}{15}$ & $7$ & $28.5333$ & $5.5047$ & $34.038$ \\	
		\hline
		\rule{0pt}{3.5ex}
		10 &\makecell{KNMD scheme with\\ centralized data placement} & ($\alpha^{\star}=\frac{5}{33}, \beta^{\star}=0$) & $7$ & $6$& $\frac{1}{14}$ & $\frac{1}{15}$ & $1.1193\times 10^{5} $ & $7.8587$ & $4.60$ & $12.4587$ \\	
		\hline
		\rule{0pt}{3.5ex}
		11 &\makecell{Scheme A in \cite{KNMD} with\\ centralized data placement} & ($\alpha=1, \beta=1$) & $7$ & $6$& $\frac{1}{14}$ & $\frac{1}{15}$ & $7 $ & $30$ & $4.60$ & $34.60$ \\	
		\hline
		\rule{0pt}{3.5ex}
		12 &\makecell{Scheme B in \cite{KNMD} with\\ centralized data placement} & ($\alpha=0, \beta=0$) & $7$ & $6$& $\frac{1}{14}$ & $\frac{1}{15}$ & $1.148\times 10^{4} $ & $10.4667$ & $4.6$ & $15.0667$ \\	
		\hline
	\end{tabular}
	\caption{Comparison of the proposed hierarchical scheme with the schemes in \cite{KNMD}}
	\label{tab:comp_HPDA}
\end{table*}
\noindent $\bullet$ Comparison with the KNMD scheme: From (\ref{eq:knmd}) it is understood that the subpacketization level of the KNMD scheme is $F_{KNMD}= max \left\{ \binom{K_1K_2}{\lfloor \frac{K_1K_2(1-\beta)M_2}{(1-\alpha)N} \rfloor } , \binom{K_1K_2}{\lceil \frac{K_1K_2(1-\beta)M_2}{(1-\alpha)N} \rceil }\right\} $, which grows exponentially with $K_1K_2$ for a given $\frac{M_2}{N}$, irrespective of the memory-regimes in (\ref{eq:knmd2}). Thus, the subpacketization level of the proposed scheme is significantly lesser than that of the KNMD scheme. For example, consider $n \ge 4$, $k=n-1$, $j=n-2$ and $i < \frac{n}{4}$ in Corollary \ref{cor:HPDA}. We obtain an $\binom{n}{i}$-division $\left(n,n-1;M_1,M_2;N \right)$ coded caching scheme with $\frac{M_1}{N}=\frac{i}{n}$, $\frac{M_2}{N}=\frac{i(n-i)}{n(n-1)}$. For the same system parameters, by (\ref{eq:knmd2}) the KNMD scheme has $\alpha^\star= \beta^\star=\frac{i}{n}$ and requires a subpacketization level, $F_{KNMD} = \binom{n(n-1)}{i(n-i)} \gg \binom{n}{i} = F_{proposed}$. A general comparison of the coding delay between the proposed scheme and the KNMD scheme is hard, as it requires determining the values of $\alpha^\star$ and $\beta^\star$ for each set of system parameters. Therefore, we consider specific examples, one from each of the three memory regimes specified in (\ref{eq:knmd2}), as presented in Table \ref{tab:comp_HPDA}. From the examples in rows $1$ and $2$, rows $5$ and $6$, and rows $9$ and $10$ of Table \ref{tab:comp_HPDA}, it is clear that irrespective of the memory regime, the proposed schemes have a significant reduction in subpacketization level compared to the KNMD scheme, at the expense of an increase in coding delay.  

\noindent $\bullet$ Comparison with the Scheme A in \cite{KNMD}: For $\alpha=1$ and $\beta=1$, the KNMD scheme reduces to the Scheme A in \cite{KNMD}. By (\ref{eq:knmd}), the loads achieved by Scheme A in \cite{KNMD} are $R_{1_A} \triangleq  K_2\cdot r\left(\frac{M_1}{N},K_1\right)$ and $ R_{2_A} \triangleq  r\left(\frac{M_2}{N},K_2\right)$. For the system parameters in the proposed scheme in Theorem \ref{thm:HPDA}, $\frac{K_1M_1}{N}=\frac{\lambda\binom{v}{t}}{\binom{k}{t}}\left(1-\frac{\binom{k}{i}}{\binom{v}{i}}\right)=\frac{\lambda\binom{v}{t}}{\binom{k}{t}}-\frac{\lambda\binom{v}{t}\binom{k}{i}}{\binom{k}{t}\binom{v}{i}}=\frac{\lambda\binom{v}{t}}{\binom{k}{t}}-\frac{\lambda \binom{v-i}{t-i}}{\binom{k-i}{t-i}}=b-\lambda_{i}$, is an integer, and $\frac{K_2M_2}{N}=\binom{k}{j}\left(\frac{\binom{k}{i}-\binom{j}{i}}{\binom{v}{i}}\right)$, need not be an integer. Therefore, the subpacketization level of the Scheme A in \cite{KNMD} is $F_{A}=max \left\{ \binom{K_1}{\frac{K_1M_1}{N}}, \binom{K_2}{\lfloor \frac{K_2M_2}{N} \rfloor } , \binom{K_2}{\lceil \frac{K_2M_2}{N} \rceil }\right\} \ge \binom{K_1}{\frac{K_1M_1}{N}} = \binom{b}{b-\lambda_{i}} =\binom{b}{\lambda_{i}}$. For $i <t$, $\binom{b}{\lambda_{i}} \ge \binom{v}{i}$ typically holds for any $t$-designs. Therefore, $F_{A} \ge \binom{b}{\lambda_{i}} \ge \binom{v}{i} = F_{proposed}$. This is illustrated with an example in rows $1$ and $3$ of Table \ref{tab:comp_HPDA}. In this example, the proposed schemes have a lesser subpacketization level as well as lesser coding delay. However, the coding delay of proposed scheme need not be lesser always.  Next, we compare the coding delay in the case where both schemes have same subpacketization levels. Consider $n>3$, $k=n-1$ and $j=n-2$ in Corollary \ref{cor:HPDA}.  We obtain an $\binom{n}{i}$-division $\left(n,n-1;M_1,M_2;N \right)$ coded caching scheme with $\frac{M_1}{N}=\frac{i}{n}$, $\frac{M_2}{N}=\frac{i(n-i)}{n(n-1)}$, $R_1=\frac{2(n-i-1)(n-i)}{(i+1)(i+2)}$ and $R_2=\frac{2n^2-3ni+i^2-2n+i+ni^2}{n(i+1)}$. For the same system parameters, Scheme A in \cite{KNMD} have subpacketization level $F_A=\binom{n}{i}$, load $R_{1_A}=\frac{(n-1)(n-i)}{i+1}$ and load $R_{2_A}$ need to be found out by memory sharing since $\frac{K_2M_2}{N}=\frac{i(n-i)}{n}$ need not be an integer. For $i=1$, $R_{2_A}=\frac{1}{n}r\left(0,n-1\right)+\frac{n-1}{n}r\left(\frac{1}{n-1},n-1\right)=\frac{n-1}{2}$. Thus, for $k=n-1$, $j=n-2$  and $i=1$, $\frac{T_{proposed}}{T_A}=\frac{R_1+R_2}{R_{1_A}+R_{2_A}}=\frac{\frac{(n-1)(n-2)}{3}+\frac{(n-1)^2}{n}}{\frac{(n-1)^2}{2}+\frac{n-1}{2}}=\frac{2(n^2+n-3)}{3n^2} < 1, \forall n> 3$. That is, for the same system parameters and subpacketization level, the proposed scheme has lesser coding delay compared to the Scheme A in \cite{KNMD}. In the examples in rows $5$ and $7$, as well as rows $9$ and $11$ of Table \ref{tab:comp_HPDA}, the proposed scheme achieves a lower coding delay than Scheme A in \cite{KNMD}, while maintaining the same subpacketization level.
\begin{table*}[!htbp]
	\centering
	\setlength{\tabcolsep}{3pt}
	\scriptsize
	\begin{tabular}{|c|c| c | c | c | c | c | c | c | c | c |}
		\hline
		\rule{0pt}{4ex}
		\makecell{Sl. \\ No.}& \makecell{Comparing Schemes}& \makecell{Parameters}  & \makecell{$K_1$} & \makecell{$K_2$}& \makecell{$\frac{M_1}{N}$} & \makecell{$\frac{M_2}{N}$} & \makecell{$F$}& \makecell{$R_1$} &\makecell{$R_2$}& \makecell{$T=$ \\ $R_1+R_2$} \\ [4pt]
		\hline
		\rule{0pt}{3.5ex}
		1 & \makecell{Proposed scheme \\ in Theorem \ref{thm:HPDA}}   & \makecell{ for $3-(8,4,1)$ design \\ with $j=2$ and $i=1$}  & $14$ & $6$ & $\frac{1}{2}$& $\frac{1}{4}$ & $8$ & $3$ & $4$ & $7$ \\
		\hline
		\rule{0pt}{3.5ex}
		2 & \makecell{Scheme II in \cite{KYWM} } & \makecell{ $\mathbf{A}$ and $\mathbf{B}$  are the optimal PDAs given\\  by Theorem 4.2 in \cite{Wei}. $\mathbf{A}=(14,2,1,7)$ \\ PDA and  $\mathbf{B}=(6,4,1,11)$ PDA }   & $14$ & $6$ & $\frac{1}{2}$& $\frac{1}{4}$ & $8$ & $9.625$ & $2.75$ & $12.375$ \\
		\hline
		\hline
		\rule{0pt}{3.5ex}
		3 & \makecell{Proposed scheme \\ in Corollary \ref{cor:HPDA} } & \makecell{  for $n=8, k=6, j=2 \text{ and } i=1$} & $28$ & $15$& $\frac{1}{4}$ & $\frac{1}{2}$ & $8$ & $15$ & $15$ & $30$ \\
		\hline
		\rule{0pt}{3.5ex}
		4 & \makecell{Scheme II in \cite{KYWM} } & \makecell{ $\mathbf{A}$ and $\mathbf{B}$ are the optimal PDAs\\ given by Theorem 4.2  in \cite{Wei} $\mathbf{A}=(28,4,1,42)$ \\ PDA, $\mathbf{B}=(15,2,1,8)$ PDA } & $28$ & $15$& $\frac{1}{4}$ & $\frac{1}{2}$ & $8$ & $42$ & $4$ & $46$ \\
		\hline
		\hline
		\rule{0pt}{3.5ex}
		5 & \makecell{Proposed scheme in \\ Corollary \ref{cor:HPDA} } &   for $n=9, k=6, j=3 \text{ and } i=1$ & $84$ & $20$& $\frac{1}{3}$ & $\frac{1}{3}$ & $9$ & $80$ & $40$ & $120$ \\
		\hline
		\rule{0pt}{3.5ex}
		6 & \makecell{Scheme II in \cite{KYWM} } & \makecell{ $\mathbf{A}$ and $\mathbf{B}$ are the optimal PDAs \\ given by Theorem 4.2 in \cite{Wei} \\ $\mathbf{A}=(84,3,1,84)$ PDA, $\mathbf{B}=(20,3,1,21)$ PDA } & $84$ & $20$& $\frac{1}{3}$ & $\frac{1}{3}$ & $9$ & $196$ & $7$ & $203$ \\
		\hline
		\hline
		\rule{0pt}{3.5ex}
		7 & \makecell{Proposed scheme in \\ Corollary \ref{cor:HPDA}  } &  for $n=8, k=7, j=4 \text{ and } i=4$ & $8$ & $35$& $\frac{1}{2}$ & $\frac{17}{35}$ & $70$ & $0.057$ & $17.55$ & $17.607$ \\
		\hline
		\rule{0pt}{3.5ex}
		8 & \makecell{Scheme II in \cite{KYWM} } & \makecell{ $\mathbf{A}=(8,2,1,4)$  being an optimal PDAs given by \\ Theorem 4.2 in \cite{Wei}  and $\mathbf{B}=(35,35,17,210)$ \\ is a PDA by Theorem 2 in\cite{Yan}  } & $8$ & $35$& $\frac{1}{2}$ & $\frac{17}{35}$ & $70$ & $12$ & $6$ & $18$ \\
		\hline
		\rule{0pt}{3.5ex}
		9 & \makecell{Scheme II in \cite{KYWM} } & \makecell{$\mathbf{A}$ and $\mathbf{B}$ are both PDAs by Theorem 2 in\cite{Yan} \\ $\mathbf{A}=(8,70,35,56)$ PDA, $\mathbf{B}=(35,35,17,210)$ PDA} & $8$ & $35$ & $\frac{1}{2}$ & $\frac{17}{35}$ & $2450$ & $4.8$ & $6$ & $10.8$ \\
		\hline
		\hline
		\rule{0pt}{3.5ex}
		10 & \makecell{Proposed scheme \\ in Corollary \ref{cor:HPDA}}  & for $n=8, k=7, j=6 \text{ and } i=4$ & $8$ & $7$& $\frac{1}{2}$ & $\frac{2}{7}$ & $70$ & $0.80$ & $4.10$ & $4.90$ \\	
		\hline
		\rule{0pt}{3.5ex}
		11 & \makecell{Scheme II in \cite{KYWM} } & \makecell{ $\mathbf{A}$ and $\mathbf{B}$ are both MAN PDAs    $\mathbf{A}=(8,70,35,56)$ \\ PDA,  $\mathbf{B}=(7,21,6,35)$ PDA } & $8$ & $7$ & $\frac{1}{2}$& $\frac{2}{7}$ & $1470$ & $1.333$ & $1.667$ & $3$ \\
		\hline
		\rule{0pt}{3.5ex}
		12 & \makecell{Scheme II in \cite{KYWM} } & \makecell{  $\mathbf{A}=(8,8,4,8)$ PDA obtained \\ by Theorem 4 in \cite{YCT} and \\ $\mathbf{B}=(7,21,6,35)$ MAN PDA  } & $8$ & $7$ & $\frac{1}{2}$& $\frac{2}{7}$ & $168$ & $1.667$ & $1.667$ & $3.333$ \\
		\hline
	\end{tabular}
	\caption{Comparison of the proposed hierarchical schemes with Scheme II in \cite{KYWM}}
	\label{tab:comp_HPDA2}
\end{table*}

\noindent $\bullet$ Comparison with the Scheme B in \cite{KNMD}: For $\alpha=0$ and $\beta=0$, the KNMD scheme reduces to the Scheme B in \cite{KNMD}. By (\ref{eq:knmd}), the loads achieved by Scheme B in \cite{KNMD} are $R_{1_B} \triangleq  r\left(\frac{M_2}{N},K_1K_2\right)$ and $ R_{2_B} \triangleq  r\left(\frac{M_2}{N},K_2\right)$, and requires a subpacketization level $F_B= max \left\{ \binom{K_1K_2}{\lfloor \frac{K_1K_2M_2}{N} \rfloor } , \binom{K_1K_2}{\lceil \frac{K_1K_2 M_2}{N} \rceil }\right\} $, which grows exponentially with $K_1K_2$ for a given $\frac{M_2}{N}$. Thus, the subpacketization level of the proposed scheme is significantly lesser than that of the Scheme B in \cite{KNMD}. For example, consider $i=j=t$ in Theorem \ref{thm:HPDA}. We obtain an $\binom{v}{t}$-division $\left(\frac{\lambda\binom{v}{t}}{\binom{k}{t}},\binom{k}{t};M_1, M_2; N \right)$ coded caching scheme with $\frac{M_1}{N}=1-\frac{\binom{k}{t}}{\binom{v}{t}}$ and $\frac{M_2}{N}=\frac{\binom{k}{t}-1}{\binom{v}{t}}$. For the same system parameters, Scheme B in \cite{KNMD} requires a subpacketization level, $F_{B} = \binom{\lambda\binom{v}{t}}{\lambda\binom{k}{t}-\lambda} \gg \binom{v}{t} = F_{proposed}$. Since $\frac{K_1K_2M_2}{N}$ and $\frac{K_2M_2}{N}$ are not always integers, a general comparison of coding delay between the proposed scheme and Scheme B in \cite{KNMD} is difficult. From the examples in Table \ref{tab:comp_HPDA}, it can be seen that the proposed scheme achieves lower coding delay in some cases (rows $1$ and $4$), while scheme B performs better in others (rows $5$ and $8$, rows $9$ and $12$). However, the proposed scheme always requires significantly lesser subpacketization.

\subsection{Comparison with  \cite{KYWM}:} Two classes of HPDAs are constructed in \cite{KYWM}. The first class of HPDAs is constructed from a given $(K,F,Z,S)$ MAN PDA and requires a subpacketization level $F=\binom{K_1K_2}{\frac{K_1K_2(M_1+M_2)}{N}}$ which grows exponentially with the total number of users $K_1K_2$ (referred to as Scheme I in \cite{KYWM}). The second class of HPDAs is constructed from two given PDAs, $\mathbf{A}=(K_1, F_1, Z_1, S_1)$ and $\mathbf{B}=(K_2, F_2, Z_2, S_2)$, resulting in an $F$-division $(K_1,K_2;M_1,M_2;N)$ coded caching scheme with memory ratios $\frac{M_1}{N}=\frac{Z_1}{F_1}$, $\frac{M_2}{N}=\frac{Z_2}{F_2}$, loads $R_1=\frac{S_1S_2}{F_1F_2}$, $R_2=\frac{S_2}{F_2}$ and subpacketization level $F=F_1F_2$ (referred to as Scheme II in \cite{KYWM}). We look for existing PDAs using which Scheme II in \cite{KYWM} gives hierarchical schemes with the same system parameters and subpacketization level as ours. As a result, a general comparison is not possible. However, illustrative examples in Table \ref{tab:comp_HPDA2} (rows $1$-$8$) show that, for schemes with the same subpacketization and system parameters, the proposed schemes achieve lower coding delay than those obtained by Scheme II in \cite{KYWM}. HPDAs constructed using two optimal PDAs in Scheme II in \cite{KYWM} exhibit higher coding delay than our schemes with the same subpacketization levels, as illustrated in rows $1$-$6$ of Table \ref{tab:comp_HPDA2}. Since the PDAs used in Scheme II in these examples are optimal PDAs, no better coding delay can be achieved using Scheme II with any other PDAs, resulting in schemes with the same system parameters and subpacketization levels. This clearly highlights the advantage of proposed schemes. Schemes with the same system parameters as ours but with different subpacketization levels can be obtained using Scheme II in \cite{KYWM} with appropriately chosen PDAs. This is illustrated in rows $7$, $9$ and $10$–$12$ of Table \ref{tab:comp_HPDA2}, and rows $1$ and $6$ of Table \ref{tab:comp_HPDA3}. In all these examples, the proposed schemes achieve a lower subpacketization level, at the expense of an increase in coding delay. 
	
\subsection{Comparison with  \cite{ZZWXL}:} Zhang \textit{et al.} in \cite{ZZWXL} proposed two centralized coded caching schemes for a hierarchical two-layer network. The first joint caching scheme (referred to as the JC scheme) with $\lambda=M_1/N$ achieves a transmission load given by, $ R_1 = K_1K_2\left(1-\frac{M_1}{N}\right)\left(1-\frac{M_2}{N}\right)\frac{1}{1+K_1M_1/N}, R_2 = K_2\left(1-\frac{M_2}{N}\right)\frac{1}{1+K_2M_2/N}$, where $M_i\in\{0,\frac{N}{K_i},\frac{2N}{K_i},\ldots,N\}$, $i=1,2$. This scheme requires a subpacketization level of $F=\binom{K_1}{K_1M_1/N}\binom{K_2}{K_2M_2/N}$. For the system parameters in the proposed scheme in Theorem \ref{thm:HPDA}, $\binom{K_1}{K_1M_1/N}=\binom{b}{\lambda_{i}}$ and for $i <t$, $\binom{b}{\lambda_{i}} \ge \binom{v}{i}$ typically holds for any $t$-designs. Therefore, for the same system parameters, the subpacketization level of the JC scheme, $F=\binom{K_1}{K_1M_1/N} \binom{K_2}{K_2M_2/N} = \binom{b}{\lambda_{i}}\binom{K_2}{K_2M_2/N} \ge \binom{v}{i}\binom{K_2}{K_2M_2/N} \gg \binom{v}{i} =F_{proposed}$. Since $\frac{K_2M_2}{N}$ of our proposed scheme is not an integer always, it is hard to have a general comparison of the coding delay. For any positive integer $q \ge 2$, consider $n=q^2$, $k=q^2-1$, $j=q^2-2$ and $i=q$ in Corollary \ref{cor:HPDA}, for which  $\frac{K_2M_2}{N}$ is an integer. For these parameters we obtain an $\binom{q^2}{q}$-division $\left(q^2,q^2-1;M_1,M_2;N \right)$ coded caching scheme with $\frac{M_1}{N}=\frac{1}{q}$, $\frac{M_2}{N}=\frac{1}{q+1}$, $R_1=\frac{2q(q^2-q-1)(q-1)}{(q+1)(q+2)}$ and $R_2=\frac{(3q^2-1)(q-1)}{q(q+1)}$. For the same system parameters, the JC scheme has subpacketization level $F_{JC}=\binom{q^2}{q}\binom{q^2-1}{q-1}$, loads $R_{1_{JC}}=\frac{q^2(q-1)^2}{q+1}$ and $R_{2_{JC}}=q-1$. Thus, $\frac{F_{JC}}{F_{proposed}}=\binom{q^2-1}{q-1} \gg 1$ and $\frac{T_{JC}}{T_{proposed}}=\frac{R_{1_{JC}}+R_{2_{JC}}}{R_1+R_2}=\frac{\frac{q^2(q-1)^2}{q+1}+q-1}{\frac{2q(q^2-q-1)(q-1)}{(q+1)(q+2)}+\frac{(3q^2-1)(q-1)}{q(q+1)}}=\frac{q^{5}+q^4-q^{3}+3q^2+2q}{2q^4+q^{3}+4q^2-q-2} > 1, \forall q$. That is, in the considered case, the proposed scheme has advantage in both subpacketization level and coding delay. The second scheme in \cite{ZZWXL} is a hybrid scheme  which requires a subpacketization level that grows exponentially with the total number of users $K_1K_2$, and therefore the proposed scheme has significantly lower subpacketization level.
\begin{table*}[!htbp]
	\centering
	\setlength{\tabcolsep}{4pt}
	\scriptsize
	\begin{tabular}{| c | c | c | c | c |}
		\hline
		\rule{0pt}{4ex}
		\makecell{Schemes and Parameters}  & \makecell{Subpacketization level \\ $F$} & \makecell{First-layer load \\$R_1$} & \makecell{Second-layer load \\$R_2$} & \makecell{Coding delay \\ $T= R_1+R_2$} \\ [4pt]
		\hline
		\rule{0pt}{3.5ex}
		\makecell{Proposed scheme in Corollary \ref{cor:HPDA}\\ $n=q^2$, $k=q^2-1$, $j=q^2-2$ and $i=q$} & $\binom{q^2}{q}$ & $\frac{2q(q^2-q-1)(q-1)}{(q+1)(q+2)}$ & $\frac{(3q^2-1)(q-1)}{q(q+1)}$ & $\frac{(q-1)(2q^4+q^3+4q^2-q-2)}{q(q+1)(q+2)}$ \\
		\hline
		\rule{0pt}{3.5ex}
		\makecell{KNMD scheme in \cite{KNMD} with centralized \\data placement, $\alpha^{\star}=\frac{1}{q}, \beta^{\star}=\frac{1}{q}$} & $\binom{q^2(q^2-1)}{q^2(q-1)}$ & $\frac{q^2(q-1)^2}{1+q^2(q-1)}$ & $q-1$ & $\frac{(q-1)(2q^3-2q^2+1)}{1+q^2(q-1)}$ \\
		\hline
		\rule{0pt}{3.5ex}
		\makecell{Scheme A in \cite{KNMD} with\\ centralized data placement} & $\binom{q^2}{q}$ & $q(q-1)^2$ & $q-1$ & $(q-1)(q^2-q+1)$ \\
		\hline
		\rule{0pt}{3.5ex}
		\makecell{Scheme B in \cite{KNMD} with\\ centralized data placement} & $\binom{q^2(q^2-1)}{q^2(q-1)}$ & $\frac{q^3(q-1)}{1+q^2(q-1)}$ & $q-1$ & $\frac{(q-1)(2q^3-q^2+1)}{1+q^2(q-1)}$ \\
		\hline
		\rule{0pt}{3.5ex}
		\makecell{Scheme I in \cite{KYWM} \\ $t=\frac{K_1K_2(M_1+M_2)}{N}=q(q-1)(2q+1)$} & $\binom{q^2(q^2-1)}{q(q-1)(2q+1)}$ & $\frac{q^2-q-1}{2q+1}$ & $\approxeq \frac{q^2-q-1}{2q+1}$ & $\approxeq \frac{2(q^2-q-1)}{2q+1}$ \\
		\hline
		\rule{0pt}{3.5ex}
		\makecell{Scheme II in \cite{KYWM} with $\mathbf{A}=(q^2,q^{q-1},q^{q-2},q^q-q^{q-1})$ PDA and \\ $\mathbf{B}=(q^2-1,(q+1)^{q-2},(q+1)^{q-3}, (q+1)^{q-1}-(q+1)^{q-2})$ \\ PDA, both obtained by Theorem 4 in \cite{YCT}} & $q^{q-1}(q+1)^{q-2}$ & $q(q-1)$ & $q$ & $q^2$ \\
		\hline
		\rule{0pt}{3.5ex}
		\makecell{JC Scheme in \cite{ZZWXL}} & $\binom{q^2}{q}\binom{q^2-1}{q-1}$ & $\frac{q^2(q-1)^2}{q+1}$ & $q-1$ & $\frac{(q-1)(q^3-q^2+q+1)}{q+1}$ \\
		\hline
	\end{tabular}
	\caption{Comparison of various hierarchical schemes with $K_1=q^2$, $K_2=q^2-1$, $\frac{M_1}{N}=\frac{1}{q}$ and $\frac{M_2}{N}=\frac{1}{q+1}$, $\forall q \ge 2$.}
	\label{tab:comp_HPDA3}
\end{table*} 
\subsection{Comparison with  \cite{WWCY}:} Wang \textit{et al.} in \cite{WWCY} proposed a centralized and a decentralized hierarchical scheme which allows a concurrent transmission between the two layers. For  all $\alpha, \beta \in[0,1]$, and memory size $M_i\in\{0,{N}/{K_i},{2N}/{K_i},\ldots,N\}$, $i=1,2$, the coding delay for the centralized hierarchical scheme in \cite{WWCY} is given by $
	T \triangleq \alpha R_\mathsf{s1,C}+(1-\alpha)R_\mathsf{s2,C}$, where $ R_\textnormal{s1,C} \triangleq R_\mathsf{p1,C} +R_\mathsf{p2,C},  R_\textnormal{s2,C} \triangleq r\left(\frac{(1-\beta) M_{2}}{(1-\alpha) N},K_{1} K_{2}\right), R_\mathsf{p1,C} \triangleq r\left(\frac{M_{1}}{\alpha N},K_{1}\right) r\left(\frac{\beta M_{2}}{\alpha N},K_{2}\right) \text{ and } R_\mathsf{p2,C} \triangleq  \min \left\{\frac{M_{1}}{\alpha N}, 1\right\}$ $r\left(\frac{\beta M_{2}}{\alpha N},K_{2}\right)$.
The approximately optimal values of $\alpha$ and $\beta$ are given in \cite{WWCY}. This scheme requires a subpacketization level $F=max \left\{ \binom{K_1K_2}{\lfloor \frac{K_1K_2(1-\beta)M_2}{(1-\alpha)N} \rfloor } , \binom{K_1K_2}{\lceil \frac{K_1K_2(1-\beta)M_2}{(1-\alpha)N} \rceil }\right\} $, which grows exponentially with the total number of users $K_1K_2$. Therefore the proposed class of HPDAs in Theorem \ref{thm:HPDA} gives schemes with significantly lower subpacketization level, at the expense of an increase in coding delay. This is illustrated in the following example.
\begin{example}
	For $n=8, k=7, j=6$ and $i=4$, the proposed scheme in Corollary \ref{cor:HPDA} gives a hierarchical scheme with $K_1=8, K_2=7, \frac{M_1}{N}=\frac{1}{2}, \frac{M_2}{N}=\frac{2}{7}, F=70$ and $T=4.90$. For the same system parameters, the centralized scheme in \cite{WWCY} has $T=1.9525$ for $\alpha^* = \beta^* = \gamma = \frac{41}{56}$ with $F=4.165\times10^{13}$. 
\end{example}

Next, we conduct a comprehensive comparison of the proposed scheme with the existing schemes discussed above, for the same system parameters, by considering a specific class of proposed HPDAs. For any positive integer $q \ge 2$, for $n=q^2$, $k=q^2-1$, $j=q^2-2$ and $i=q$ in Corollary \ref{cor:HPDA}, we obtain a hierarchical coded caching scheme with $K_1=q^2$, $K_2=q^2-1$, $\frac{M_1}{N}=\frac{1}{q}$ and $\frac{M_2}{N}=\frac{1}{q+1}$. For the same system parameters, the required subpacketization level and achieved coding delay of the proposed scheme and the schemes  in \cite{KNMD}, \cite{KYWM} and \cite{ZZWXL} are given in Table \ref{tab:comp_HPDA3}. It can be verified from Table \ref{tab:comp_HPDA3} that for all $q \ge 5$, the subpacketization levels of various schemes  satisfy the order $F_{proposed}=F_{scheme A in [8]} < F_{scheme II in [9]} < F_{JC} < F_{KNMD}=F_{scheme B in [8]} < F_{scheme I in [9]}$. Similarly, the coding delay follows the order $T_{scheme I in [9]} < T_{KNMD} < T_{scheme B in [8]} < T_{scheme II in [9]} < T_{proposed} < T_{JC} < T_{scheme A in [8]}$. The proposed scheme has advantage in coding delay with the same subpacketization level compared to Scheme A in \cite{KNMD}, has advantage in both subpacketization and coding delay compared to the JC scheme, and has significantly lesser subpacketization level compared to the other schemes at the expense of an increase in coding delay. 

The above comparisons have shown that the proposed hierarchical schemes in Theorem \ref{thm:pda} have significantly lower subpacketization levels compared to many known hierarchical schemes. In cases where the system parameters and the subpacketization level of the proposed scheme and the existing scheme match, the proposed scheme has a better coding delay. 
     \section{Performance analysis of the proposed class of PDAs}\label{perform_anlysis2}
     In \cite{RS}, two classes of PDAs are constructed from a given $t-(v,k,1)$ design. The class of PDAs in Theorem \ref{thm:pda} subsumes as special cases these two classes of PDAs, as shown below.  
     \begin{itemize}[left=0pt]
     	\item For $\lambda=1$ and $i=1$, Theorem \ref{thm:pda} gives a $(v-j+1)-\left( \frac{\binom{v}{t}\binom{k}{j}}{\binom{k}{t}},v,v-j,\frac{\binom{v}{j-1}\lambda\binom{v-j}{t-j}}{\binom{k-j}{t-j}} \right)$ PDA, for any $j \in [t]$. This is exactly the class of PDAs in Scheme I in \cite{RS}.
     	\item For $\lambda=1$ and $j=i+1$, Theorem \ref{thm:pda} gives a $\binom{v-1}{i}-\left( \frac{\binom{v}{t}\binom{k}{i+1}}{\binom{k}{t}},\binom{v}{i},\binom{v}{i}-(i+1),\frac{v\binom{v-(i+1)}{t-(i+1)}}{\binom{k-(i+1)}{t-(i+1)}} \right)$ PDA, $\forall i \in [t-1]$. This is exactly the class of PDAs in Scheme II in \cite{RS}. 
     \end{itemize}
    
     Now we compare the class of PDAs in Corollary \ref{cor:HPDA} and some existing PDAs with the same system parameters. The class of PDAs constructed either subsumes several known PDAs or achieves a better load for the same system parameters. 

	\noindent $\bullet$ Comparison with \cite{YCT}: For $k=n$ and $j=n-1$, Corollary \ref{cor:HPDA} gives a $\left(\binom{n}{n-1}, \binom{n}{i}, \binom{n}{i}-\binom{n-1}{i}, \right.$ $\left. \binom{n}{n-1-i}\right) \equiv \left(n, \binom{n}{i}, \binom{n-1}{i-1}, \binom{n}{i+1}\right)$ PDA $ \forall i < n $. This is exactly the MAN PDA in \cite{YCT}, which is the PDA representation of the MAN scheme in \cite{MaN}, for all non trivial values of $i$.
	
	\noindent $\bullet$ Comparison with \cite{SCYG} :Two class of PDAs are constructed using a 3-uniform 3-partite hypergraph in \cite{SCYG}. For $k=n, j=n-a$ and $i=b$, where $a,b \in \mathbb{Z}^+$, Corollary \ref{cor:HPDA} gives a $\left(\binom{n}{n-a}, \binom{n}{b}, \binom{n}{b}-\binom{n-a}{b}, \binom{n}{n-a-b}\right)$ $\equiv \left(\binom{n}{a}, \binom{n}{b}, \binom{n}{b}-\binom{n-a}{b}, \binom{n}{a+b}\right)$ PDA, where $i \leq j$. Therefore $i \leq j \implies b \leq n-a \implies a+b \leq n$.  This is exactly the first class of PDAs constructed in \cite{SCYG}.
	
\begin{figure}[!htbp]
	\centering
	\captionsetup{justification=centering}
	\includegraphics[width=0.45\textwidth]{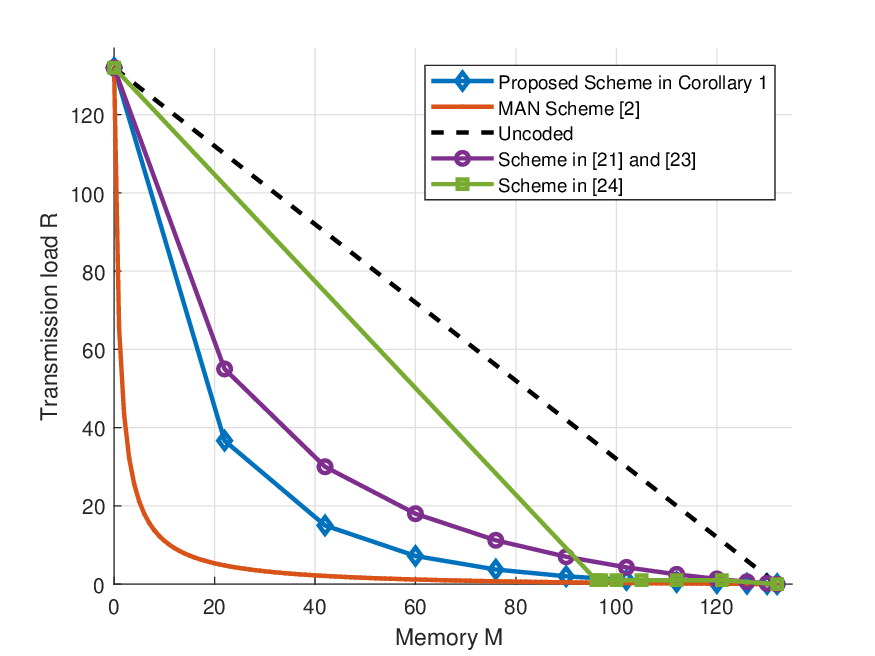}
	\caption{The transmission loads as a function of memory of schemes in Corollary \ref{cor:HPDA}, \cite{MaN}, \cite{JeQi}, \cite{ZCJ} and \cite{MWZW}.}
	\label{fig:pdacomp}
	\end{figure}

	\noindent $\bullet$ Comparison with \cite{JeQi} : In \cite{JeQi}, for $a, b, n, t \in \mathbb{Z}^+$ such that $a+b \leq n$ and $0 \leq t < b$, a  $\left(\binom{n}{a+t}\binom{a+t}{a}, \binom{n}{b-t}, \right.$ $\left. \binom{n}{b-t}-\binom{n-a-t}{b-t}, \binom{n}{a+b}\binom{a+b}{b} \right) $ PDA is constructed using strong edge coloring of a bipartite graph via combining two strong edge colorings of bipartite graphs that have the same set of colors. For $k=n-a, j=n-a-t$ and $i=b-t$, where $a,b,t \in \mathbb{Z}^+$, Corollary \ref{cor:HPDA} gives a $\left(\binom{n}{n-a}\binom{n-a}{n-a-t}, \binom{n}{b-t}, \binom{n}{b-t}-\binom{n-a-t}{b-t}, \binom{n}{n-a-b}\binom{a+t}{t} \right)$ $ \equiv \left(\binom{n}{a+t}\binom{a+t}{a}, \binom{n}{b-t}, \binom{n}{b-t}-\binom{n-a-t}{b-t}, \right.$ $ \left. \binom{n}{a+b}\binom{a+t}{t} \right) $ PDA. Since $i \leq j$, $i \leq j \implies b-t \leq n-a-t \implies a+b \leq n$. Since $t < b$, the proposed class of PDAs in Theorem \ref{thm:pda} has exactly the same $K, F \text{ and } Z$ as that of the scheme in \cite{JeQi}, but a smaller $S$. Thus proposed PDAs result in reduced transmission load for the same system parameters. This is illustrated with an example in Fig.\ref{fig:pdacomp}.

	\noindent $\bullet$ Comparison with \cite{MJXQ}: For $k=n$ and $j=n-1$, the transpose of $\mathbf{P}$ gives a $\left(\binom{n}{i}, n, i, \binom{n}{i+1}\right)$ PDA $ \forall i < n $. This is the First Variant MAN PDA in \cite{MJXQ}. For $k=n$ and $j=k+1$, Corollary \ref{cor:HPDA} gives a $\left(\binom{n}{i+1}, \binom{n}{i}, \binom{n}{i}-(i+1), n \right) $ PDA $ \forall i < n $. This is the Second Variant MAN PDA in \cite{MJXQ}.
	
	\noindent $\bullet$ Comparison with \cite{ZCJ}: For $a, b, n, t \in \mathbb{Z}^+$ such that $a+b \leq n$ and $ 0 \leq t < b$, \cite{ZCJ} gives a $\left(\binom{n}{a+t}\binom{a+t}{a},\binom{n}{a+b},\right.$ $ \left.\binom{n}{a+b}-\binom{n-a-t}{b-t},\binom{n}{b}\binom{b}{b-t} \right)$ PDA using concatenating constructions. The PDAs obtained when $a+b <n$, are either subsumed by the class of PDAs in Corollary \ref{cor:HPDA} or for the same values of $K,F$ and $Z$, Corollary \ref{cor:HPDA} gives PDAs with smaller values of $S$. Apart from comparable ones, Corollary \ref{cor:HPDA} gives PDAs that are not obtained by the scheme in \cite{ZCJ}. For $k=n-a, j=n-a-t, i=n-a-b$, where $a,b \in \mathbb{Z}^+$, Corollary \ref{cor:HPDA} gives a  $\left(\binom{n}{n-a}\binom{n-a}{n-a-t}, \binom{n}{n-a-b}, \binom{n}{n-a-b}-\binom{n-a-t}{n-a-b},\binom{n}{b-t}\binom{a+t}{t}\right)$  $\equiv  \left(\binom{n}{a+t}\binom{a+t}{a}, \binom{n}{a+b}, \binom{n}{a+b}-\binom{n-a-t}{b-t},\right.$ $\left.\binom{n}{b}\frac{(a+t)!}{(n-b-t)!a!}\right)$ PDA. Since $i=n-a-b, a+b=n-i \implies a+b \leq n$. Since  $i \leq j$, $i < j \implies n-a-b < n-a-t \implies t < b$. That is, in this case, for the same $K, F$ and $Z$, PDA in Corollary \ref{cor:HPDA} has lower $S$ and therefore lower load, compared to the PDA in \cite{ZCJ}. This is illustrated with an example in Fig.\ref{fig:pdacomp}.

	\noindent $\bullet$ Comparison with \cite{MWZW}:  For $m, s, t \in \mathbb{Z}^+, \omega \in \mathbb{Z}$ with $0 \leq \omega \leq t \leq s \text{ and } s+t-2\omega \leq m$, a $\left(\binom{t}{\omega}\binom{m}{t}, \binom{m}{s}, \binom{m}{s}-\binom{m-t}{s-\omega},\right.$ $ \left. \binom{m}{s+t-2\omega} \right)$ PDA is given in \cite{MWZW}. This class of PDAs subsumes the MAN PDA and the PDA given in  \cite{SCYG} discussed above, which are also subsumed by our proposed PDA in Corollary \ref{cor:HPDA}. For $m-t = s-\omega $, \cite{MWZW} gives a $\left(\binom{t}{\omega}\binom{m}{t}, \binom{m}{s}, \binom{m}{s}-1, \binom{m}{m-\omega} \right)$ $\equiv \left(\binom{m}{s}\binom{s}{\omega}, \binom{m}{s}, \binom{m}{s}-1, \binom{m}{\omega} \right)$ PDA. For $ i= j < k < n, n-j=s, \text{ and } m-j=\omega$, Corollary \ref{cor:HPDA} gives a $\left(\binom{n}{s}\binom{s}{\omega}, \binom{n}{s}, \binom{n}{s}-1, \binom{s}{\omega} \right)$ PDA. In this case, for same $K, F$ and $Z$, the PDA in Theorem \ref{thm:pda} has lower $S$ and therefore lower load. Both constructions give PDAs that are not obtained by the other. For example,  \cite{MWZW} gives a $(12,6,4,6)$ PDA for $(m,s,t,\lambda) = (4,2,2,1)$, which is not obtainable from Corollary \ref{cor:HPDA}. Conversely, Corollary \ref{cor:HPDA} gives a $(12,4,2,8)$ PDA for $(n,k,j,i) = (4,3,2,1)$, which is not obtainable from \cite{MWZW}.    	  
	\begin{figure}[!htbp]
	\centering
	\captionsetup{justification=centering}
	\includegraphics[width=0.45\textwidth]{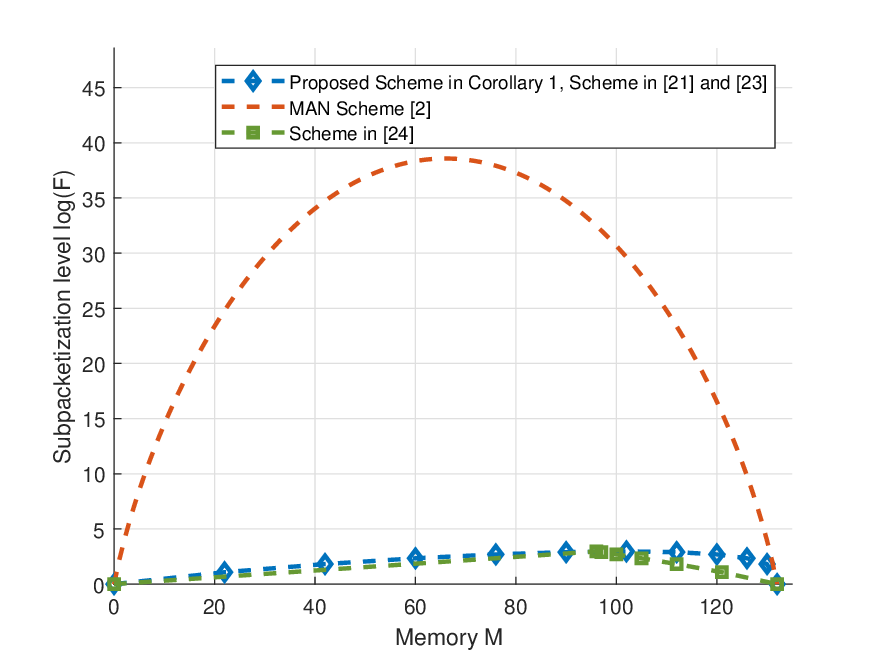}
	\caption{The subpacketization levels as a function of memory of schemes in Corollary \ref{cor:HPDA}, \cite{MaN}, \cite{JeQi}, \cite{ZCJ} and \cite{MWZW}.}
	\label{fig:subpackcomp}
	\end{figure}	
	
By choosing the parameters $n=12, k=11, j=10$ and $i=1 \text{ to } 10$ of the proposed scheme in Corollary \ref{cor:HPDA}, $K=132$ of the MAN scheme in \cite{MaN}, the parameters $n=12, a=t=1$ and $b=2 \text{ to } 11$ of the scheme in \cite{JeQi}, the parameters $n=12, a=t=1$ and $b=2 \text{ to } 10$ of the scheme in \cite{ZCJ}, the parameters $n=12, t=2, \omega=1$ and $s=2 \text{ to } 11$ of the scheme in \cite{MWZW}, coded caching schemes are obtained for $K=132$ and $N=132$, at various memory points. The load and subpacketization level of these schemes at various memory points are illustrated in Fig.\ref{fig:pdacomp} and Fig.\ref{fig:subpackcomp} respectively. It is clear that the proposed PDAs in Corollary \ref{cor:HPDA} have the advantage of subpacketization compared to the MAN scheme and have an advantage on load compared to schemes in \cite{JeQi}, \cite{ZCJ} and \cite{MWZW}.
	
\section{Conclusion}\label{concl}
	In this work, we first constructed a class of PDAs using combinatorial $t$-designs, that either subsume several known PDAs or achieve better load for the same system parameters. From this class of PDAs, we then constructed a novel class of HPDAs that give low subpacketization level hierarchical coded caching schemes. Compared with known hierarchical schemes, in cases where the subpacketization level and the system parameters are the same, the proposed scheme has a better coding delay. One possible direction for further research is to obtain low subpacketization level schemes for any given two-layer $(K_1, K_2; M_1, M_2; N)$ hierarchical caching system. Obtaining lower bounds for $R_1$ and $R_2$ under subpacketization constraint is also an open problem. 
	\section*{Acknowledgement}
	This work was supported partly by the Science and Engineering Research Board (SERB) of Department of Science and Technology (DST), Government of India, through J.C Bose National Fellowship to B. Sundar Rajan.	
	\begin{appendices}
		\section{Proof of Theorem \ref{thm:pda}}\label{appendix:PDA_proof}
		For the sake of convenience, we continue with the representation of $(Y\textbackslash X)_\alpha$ for the non-star entries.
		
		By $(\ref{eq:cons2})$, for a given column $(A,Y)$, a $\star$ appears in a row $X \in \binom{\mathcal{X}}{i}$  if and only if $X \not\subset Y $. Number of $i$ sized subset $X$ of $j$ sized set $Y$ is equal to $\binom{j}{i}$. Therefore, the number of $\star$s in a column, $Z= \binom{v}{i}-\binom{j}{i}$. Thus $C1$ of PDA definition holds.  $C2$ is obvious by construction. 
		
		Assume that $(Y\textbackslash X)_\alpha$ appears in the  column $(A,Y)$, then by $(\ref{eq:cons2})$, $X \subset Y $ and $|Y\textbackslash X|=(j-i)$. Therefore, $(Y\textbackslash X)$ will be different for each $X$. Thus   $(Y\textbackslash X)_\alpha$ appears only once in a given column. Since $\alpha$ denotes the $\alpha^{th}$ occurrence of $Y\textbackslash X$ from left to right in the row indexed by $X$, $(Y\textbackslash X)_\alpha$ cannot appear more than once in a row. Thus $C3.(a)$ of PDA definition holds. 
		
		For any two distinct entries $P_{X_i,(A_i,Y_i)}$, $P_{X_j,(A_j,Y_j)}$ with $X_i \neq X_j$ and $(A_i,Y_i) \neq (A_j,Y_j)$, let $ P_{X_i,(A_i,Y_i)} = P_{X_j,(A_j,Y_j)} = (Y\textbackslash X)_\alpha $. Then by $(\ref{eq:cons2})$, $Y_i\textbackslash X_i= Y_j\textbackslash X_j= Y\textbackslash X$. Also $ X_i\subseteq Y_i$  and $X_j \subseteq Y_j $. Therefore, $(Y_i \textbackslash X_i) \cup X_j = Y_j $. Since $X_i \neq X_j $, there exist at least one $x$ such that $x \in X_i \textbackslash X_j$, which implies $x \notin Y_i \textbackslash X_i$ and $x \notin X_j$. Therefore,  $x \notin \{(Y_i \textbackslash X_i) \cup X_j\}$ which implies $x \notin Y_j$. That means  $X_i \not\subset Y_j $. Therefore, by $(\ref{eq:cons2})$, $P_{X_i,(A_j,Y_j)}= \star $. Similarly, $P_{X_j,(A_i,Y_i)}= \star $. Thus $C3.(b)$ of PDA definition holds.
		
		The regularity \textit{g} of the above PDA is obtained by finding in how many rows $(Y\textbackslash X)$ appears. By  $(\ref{eq:cons2})$, each $(Y\textbackslash X)$ will appear in those rows $X \in \binom{\mathcal{X}}{i}$ in which  $(Y\textbackslash X)$ does not belong. Also $|Y\textbackslash X|=(j-i)$. Therefore,  $g=\binom{v-(j-i)}{i}=\binom{v-j+i}{i}$.  Thus $C2'$ of g-PDA definition holds. 
		Therefore, $\mathbf{P}$ is a $\binom{v-j+i}{i}-\left( \frac{\lambda\binom{v}{t}\binom{k}{j}}{\binom{k}{t}},\binom{v}{i},\binom{v}{i}-\binom{j}{i},\frac{\binom{v}{j-i}\lambda\binom{v-j}{t-j}}{\binom{k-j}{t-j}} \right)$ PDA. \hfill $\blacksquare$ 

\section{Proof of Theorem \ref{thm:HPDA}}\label{appendix:hpda}
\subsubsection*{HPDA Properties Verification}
By $(\ref{eq:Mirror})$, $Q_{X,k_1}^{(0)}$ is a $ \star $ if the row $X$ of $\mathbf{P^{(k_1)}}$ is a star row. The row $X$ of $\mathbf{P^{(k_1)}}$ is a star row if and only if $X \not\subset A$. Since $|A|=k$ and $|X|=i$, the number of star rows in each  $\mathbf{P^{(k_1)}}$ is $\binom{v}{i}-\binom{k}{i}$. Therefore, each column of $\mathbf{Q^{(0)}}$ has $Z_1=\binom{v}{i}-\binom{k}{i}$ stars. Thus $B1$ of HPDA definition holds.

Each $\mathbf{Q^{(k_1)}}$ is constructed by replacing the star rows of $\mathbf{P^{(k_1)}}$ by distinct integers. Since $\mathbf{P}$ is a PDA by Theorem \ref{thm:pda}, each $\mathbf{P^{(k_1)}}$ is a PDA with the number of columns equal to the number of $j$ sized subsets of a $k$ sized block $A$. Therefore, $K_2=\binom{k}{j}$. By $(\ref{eq:cons2})$, the number of stars in a given column, excluding the stars in a star row of $\mathbf{P^{(k_1)}}$ is, $Z_2 = Z-Z_1 =\left(\binom{v}{i}-\binom{j}{i}\right)-\left(\binom{v}{i}-\binom{k}{i}\right)=\binom{k}{i}-\binom{j}{i}$. 
Each $\mathbf{Q^{(k_1)}}$ is constructed by replacing the star rows of $\mathbf{P^{(k_1)}}$ by distinct integers which has no intersection with $[S]$. Let $S_{k1}$ denote the integer set of $\mathbf{Q^{(k_1)}}$. Therefore, $\mathbf{Q^{(k_1)}}$ is a $\left(\binom{k}{j},\binom{v}{i},\binom{k}{i}-\binom{j}{i},|S_{k1}|\right)$ PDA. Thus, $B2$ of HPDA definition holds.

$S_m$ denotes the set of distinct integers replacing the star entries in the star rows of $\mathbf{P^{(k_1)}}$ and $q_{X,k_1}^{(0)}$ is a $ \star $ if the row $X$ of $\mathbf{P^{(k_1)}}$ is a star row.  Thus, $B3$ of HPDA definition holds.

Let $q_{X,k_2}^{(k_1)}=q_{X',k'_2}^{(k'_1)}=s$, where $k_1 \neq k'_1$. Then $s \notin S_m$ since each integer of $S_m$ occurs exactly once. If $q_{X',k_2}^{(k_1)}=s'$ is an integer, then $s' \in S_m$ since  $\mathbf{P}$ is a PDA. If $q_{X',k_2}^{(k_1)}=s' \in S_m$, then by our construction 	$q_{X',k_1}^{(0)}=\star$. Thus, $B4$ of HPDA definition holds. 

From the construction in $(\ref{eq:cons2})$, it is clear that any sub-array $\mathbf{P^{(k_1)}}$ contains at most $\frac{\binom{k}{j-i}\lambda\binom{v-j}{t-j}}{\binom{k-j}{t-j}}$ distinct integers. Let $S'_{k1}$ denote this set of integers in $\mathbf{Q^{(k_1)}}$. Therefore, $S'_{k1}\leq \frac{\binom{k}{j-i}\lambda\binom{v-j}{t-j}}{\binom{k-j}{t-j}}$.  Let $S^{''}_{k1}$ denote the set of distinct integers in $\mathbf{Q^{(k_1)}}$ replacing the star rows of $\mathbf{P^{(k_1)}}$. There are  $Z_1$ star rows in $\mathbf{P^{(k_1)}}$ and each star row contain $K_2$ stars. Therefore, $ S^{''}_{k1}=\left[S+1+(k_1-1)Z_1K_2: S + k_1Z_1K_2\right]=$ \small $\left[S+1+(k_1-1)\left( \binom{v}{i}-\binom{k}{i} \right)\binom{k}{j}:S + k_1\left( \binom{v}{i}-\binom{k}{i} \right) \binom{k}{j}\right]$. \normalsize Thus the integer set of $\mathbf{Q^{(k_1)}}$ is given by, $
S_{k1} = S'_{k1} + S^{''}_{k1} \leq \frac{\binom{k}{j-i}\lambda\binom{v-j}{t-j}}{\binom{k-j}{t-j}} + $  \small $\left[S+1+(k_1-1)\left( \binom{v}{i}-\binom{k}{i} \right)\binom{k}{j}:S + k_1\left( \binom{v}{i}-\binom{k}{i} \right)\binom{k}{j}\right]$. \normalsize
And, $S_m=\left[ S+1 : S+K_1Z_1K_2\right] =\left[S+1 : S+\binom{v}{k}\binom{k}{j}\left(\binom{v}{i}-\binom{k}{i}\right)\right]$. The set of integers $\underset{k_1=1}{\bigcup^{K_1} S_{k_1}}\textbackslash S_m$ is same as the set of integers $[S]$ in the original PDA $\mathbf{P}$. Thus, the array $\mathbf{Q}$ is a $\left(\frac{\lambda\binom{v}{t}}{\binom{k}{t}},\binom{k}{j};\binom{v}{i}; \right. $ $\left. \binom{v}{i}-\binom{k}{i},\binom{k}{i}-\binom{j}{i};S_m, S_1,..,S_{K_1} \right)$ HPDA. \hfill $\blacksquare$ 
\end{appendices}

\end{document}